\documentclass[]{spie}  %>>> use for US letter paper
\usepackage[]{graphicx}
\usepackage{mathptm}

\def\apjl{ApJ}
\def\apjs{ApJS}
\def\apj{ApJ}
\def\aj{AJ}
\def\mnras{MNRAS}
\def\nat{Nature}
\def\araa{ARA\&A}
\def\pasp{PASP}
\def\aap{A\&Ap}
\def\ao{Applied Optics}

\def\prd{Phys.\ Rev. D}
\def\sovast{Soviet Ast.}

\def\ruo2{RuO$_{2}$}

\def\sqdeg{\ensuremath{\mathrm{deg}^2}}

\def\mum{\ensuremath{\mu \mathrm{m}}}

\title{The South Pole Telescope} 

\author{
J.\ E.\ Ruhl\supit{a},
P.\ A.\ R.\ Ade\supit{b},
J.\ E.\ Carlstrom\supit{c},
H. M.\ Cho\supit{d},
T.\ Crawford\supit{c},
M.\ Dobbs\supit{e},
C.\ H.\ Greer\supit{c},
N.\ W.\ Halverson\supit{d},
W.\ L.\ Holzapfel\supit{d},
T.\ M.\ Lanting\supit{d},
A.\ T.\ Lee\supit{d,e},
J.\ Leong\supit{a},
E.\ M.\ Leitch\supit{c},
W.\ Lu\supit{a},
M.\ Lueker\supit{d},
J.\ Mehl\supit{d},
S.\ S.\ Meyer\supit{c},
J.\ J.\ Mohr\supit{f},
S.\ Padin \supit{c},
T.\ Plagge\supit{d},
C.\ Pryke\supit{c},
D.\ Schwan\supit{d},
M.\ K.\ Sharp\supit{c},
M.\ C.\ Runyan\supit{c},
H.\ Spieler\supit{e},
Z.\ Staniszewski\supit{a} and
A.\ A.\ Stark\supit{g}
\skiplinehalf
\supit{a}Department of Physics, Case Western Reserve University, Cleveland, OH, U.S.A.;\\
\supit{b}Department of Physics and Astronomy, Cardiff University, Wales, UK;\\
\supit{c}Kavli Institute for Cosmological Physics, University of Chicago, Chicago, IL, U.S.A;\\
\supit{d}Department of Physics, University of California, Berkeley, CA, U.S.A;\\
\supit{e}Physics Division, Lawrence Berkeley National Laboratory, Berkeley, CA, U.S.A.;\\
\supit{f}Department of Astronomy, University of Illinois, Urbana-Champaign, IL, U.S.A.;\\
\supit{g}Harvard-Smithsonian Center for Astrophysics, Cambridge, MA, U.S.A.
}

\authorinfo{Send correspondence to J. Ruhl:
E-mail: ruhl@case.edu, Telephone: 216 368 4049}
\begin{document} 
  \maketitle 

%%%%%%%%%%%%%%%%%%%%%%%%%%%%%%%%%%%%%%%%%%%%%%%%%%%%%%%%%%%%% 
\begin{abstract}
A new 10 meter diameter telescope is being constructed for deployment
at the NSF South Pole research station.  The telescope is designed for
conducting large-area millimeter and sub-millimeter wave surveys 
of faint, low contrast emission, as required to map
primary and secondary anisotropies in the 
cosmic microwave background.
To achieve the required 
sensitivity and resolution, the telescope design employs an off-axis primary with a
10~m diameter clear aperture. The full aperture and the associated optics will
have a combined surface accuracy of better than 20 microns rms to
allow precision operation in the submillimeter atmospheric
windows. The telescope will be surrounded with a large reflecting
ground screen to reduce sensitivity to thermal emission from the
ground and local interference.  The optics of the telescope will
support a square
degree field of view at 2mm wavelength and will feed
a new 1000-element micro-lithographed
planar bolometric array with superconducting transition-edge sensors
and frequency-multiplexed readouts.  The first key project will
be to conduct a survey over $\sim 4000$ degrees for galaxy clusters
using the Sunyaev-Zel'dovich Effect.  This survey should find many
thousands of clusters with a mass selection criteria that is
remarkably uniform with redshift.
Armed with
redshifts obtained from optical and infrared follow-up observations, it is expected that the survey
will enable significant constraints to be placed on the equation of state
of the dark energy.

%{\bf This paper can be up to 15 pages long}
\end{abstract}

%>>>> Include a list of keywords after the abstract 


%%%%%%%%%%%%%%%%%%%%%%%%%%%%%%%%%%%%%%%%%%%%%%%%%%%%%%%%%%%%%
\section{INTRODUCTION}
\label{sec:intro}  % \label{} allows reference to this section

Remarkable progress has been made in the characterization of
the cosmic microwave background radiation (CMB) over the last several years. 
It was nearly 30 years after the
initial discovery of the CMB by Penzias and Wilson in 1965\cite{penzias65}
before small differences in its intensity were
measured by COBE\cite{smoot92,bennett96} and its spectrum was
shown to be a blackbody to high precision\cite{mather94,fixsen96}.
The remarkable isotropy, precise to a part in  $10^5$, helped motivate the
inflation theory for the origin of the universe. 
In the past few years, subsequent measurements of the first acoustic peak and 
its harmonics
in the angular power spectrum\cite{miller99,mauskopf00a,debernardis00,hanany00,halverson02,netterfield02}
provided further support for inflation by showing the curvature of the 
universe was flat. They also allowed a full accounting for the
matter-energy densities of the universe, finding in agreement with
the analysis of Type 1a supernovae observations\cite{perlmutter99,reiss98} 
that the universe is 
now dominated
by some sort of ``dark energy'' that apparently is causing
the expansion of the universe to accelerate. More recently the 
{\sl WMAP} satellite has produced spectacular all sky maps of the 
temperature anisotropy yielding a highly precise measurement of the angular
power spectrum up to multipoles of $\ell \sim 600$, corresponding to
an angular scale of $\sim 20'$\cite{bennett03}. The {\sl WMAP} data, especially combined
with finer angular scale CMB anisotropy measurements made with ACBAR\cite{kuo04} and CBI\cite{mason02}
and with other probes of large scale structure 
have provided a high degree of confidence in the now standard cosmological model
and allowed tight constraints to be placed on many of its parameters\cite{spergel03}.

While these measurements have led to rapid progress in our
understanding of the universe, they have
raised even more profound questions about the nature of
dark energy and of the possibility of directly testing inflation 
and determining its energy scale. 
Remarkably, these questions can be addressed
through future measurements of the CMB temperature
anisotropy on fine angular scales and of the CMB polarization anisotropy 
on all angular scales; they form the basis of the scientific case for the
South Pole Telescope (SPT) program.

Finer angular scale temperature anisotropy measurements are needed
to precisely measure the angular power spectrum through the damping
tail\cite{hu97}.  Such observations will lead to better parameter constraints and
in particular allow a better characterization of 
the underlying primordial matter power spectrum, 
that in principle can be used to constrain inflationary models\cite{spergel03}. 
On angular scales of a few arcminutes and smaller, i.e., multipoles
exceeding $\sim 2000$, the CMB anisotropy is dominated 
by secondary effects caused by distortions of the CMB as it passes 
through the universe. 
The largest such effect is the Sunyaev-Zeldovich
Effect\cite{sunyaev70,sunyaev72} (SZE), in which the CMB photons
are inverse Compton scattered by the hot intracluster
gas of galaxy clusters. The SZE is a potentially powerful
probe of cosmology\cite{carlstrom02}.  Perhaps its most
powerful use will be to enable large area, redshift independent
surveys for galaxy clusters. As the growth of massive clusters
is critically dependent on the underlying cosmology, the
yields from such surveys can be used to set tight constraints
on cosmological parameters and to investigate the nature
of dark energy, i.e., by determining its equation of state\cite{holder01b,haiman01}.

Measurements of the polarization of the CMB are extremely
challenging, but also have enormous potential for discovery. 
The intrinsic polarization of the CMB reflects the local
radiation field anisotropy, specifically the local quadrupole 
moment of the incident radiation field, at the surface of
last scattering 14 billion years ago\cite{hu_w97}.  The dominant contribution 
is due to Doppler shifted radiation fields arising from the acoustic oscillations
at the time of last scattering.  This causes the so called E-mode polarization
(curl free
polarization patterns on the sky)
and has been detected by DASI\cite{kovac02}, while the temperature-polarization
cross power spectrum (TE) has been detected by DASI 
and {\sl WMAP}\cite{kogut03a}\footnote{At large angular scales corresponding 
to the first few multipoles, the
detected {\sl WMAP} TE signal is anomalously strong 
indicating an early reionization of the universe\cite{kogut03a}.}.
However, if inflation 
occurred in the early universe 
at a high-enough energy scale, a portion of the
local quadrupole at last scattering will be due to primordial gravitational waves
created during the inflationary epoch\cite{polnarev85,crittenden93}.  
The inflationary gravitational
waves  will cause
both E-mode and B-mode (curl component) polarization 
patterns in the CMB\cite{seljak97a,kamionkowski97b,seljak97}. While
the B-mode pattern can be distinguished from the intrinsic E-mode
pattern, the gravitational lensing of the CMB by large scale
structure in the universe will create B-mode polarization 
from the intrinsic E-mode signal at a level that is higher
than the inflationary B-modes for all but the most optimistic inflationary
models.  In this case, the only hope in recovering the inflationary B-modes
from the lensing B-mode foreground lies in exploiting the 
different angular power spectra
and their correlations with the temperature and E-mode
spectra\cite{okamoto02,knox02}.  The lensing polarization signal is
also interesting in its own right as it can be used to trace the growth of
large scale structure which in turn is sensitive to the 
mass of the neutrino and the equation of state of the dark energy. 

This paper presents the design for a new telescope, the South Pole
Telescope (SPT), that is being designed to pursue the 
next generation CMB temperature and
polarizaton studies at the exceptional South Pole site.  The telescope
is designed explicitly for conducting large area, high sensitivity survey
observations of the temperature and polarization of the CMB.
The SPT
has an off-axis 10 meter diameter aperture to provide 1$'$ resolution
at 2mm wavelength with exceptionally low spillover.  The optics will
support a one degree diameter field of view.  To further
reduce signals due to scattering and spillover, the entire telescope
will be deployed within a large reflecting ground screen. 
Deployment is planned for late 2006, with 
first observations starting early 2007. The properties of
the South Pole
site are reviewed briefly in Section~\ref{sec:site} and the telescope
is discussed in Section~\ref{sec:telescope}.

State of the art bolometer detectors used for ground-based 
CMB observations 
are essentially background limited.  In this case, the obvious
path to higher sensitivity is to add more detectors.
The initial SPT
receiver will be dedicated for fine angular scale CMB temperature and SZE
survey observations.  It will consist of an array of 1000 micro-lithographed
bolometers with transition edge sensors (TES) read out
with a novel frequency multiplexing scheme. While a polarization-sensitive
receiver is planned, only technology development funds have been
obtained for it at this time. 
The details and
status of the SZE survey receiver are outlined in Section~\ref{sec:receiver}. 

A review of the initial SPT science goals and consideration of
astronomical foregrounds, atmospheric emission and the implications
of these on 
the definition of the receiver bands 
and the observing strategy
is given Section~\ref{sec:science}. 

\section{Site} 
\label{sec:site}

Ground based astronomical observations at mm and sub-mm 
wavelengths place extreme requirements 
on the transmission and stability of the atmosphere.  Steady
atmospheric emission loads the bolometric detectors and 
adds photon noise, reducing overall sensitivity.  Fluctuations
in atmospheric brightness add noise to the detector timestream data as well.
For these reasons high altitude sites, preferably with stable atmospheric
conditions, are needed for ground based observations such as those
targeted by the SPT.

Emission from the atmosphere consists of two components, one due to
``dry air" which results from the wings of oxygen lines and a
second component due to water vapor.
The ``dry air" component is well-mixed in the atmosphere and produces 
a signal that is only a function of elevation, commonly 
removed by beam switching.  
Water vapor, on the other hand, exhibits considerable variations in its density.
These fluctuations in water vapor density result in spatial variations
in the brightness of the sky.

The South Pole lies on the Antarctic Plateau, at an altitude of
2800~m.  
The average atmospheric pressure in the winter is 675 millibars.
In addition, the low temperature at the 
South Pole reduces the water vapor content of the atmosphere, lowering
both atmospheric emission and fluctuations in brightness.  
There is a long history of measuring atmospheric properties relevant
to millimeter-wave observing at the South Pole;
profiles of the atmospheric temperature, pressure, and water vapor
have been measured at least once a day for several decades 
by the South Pole meteorology office, using balloon-borne radiosondes.
Schwerdtfeger\cite{schwerdtfeger} comprehensively reviewed the climate 
of the Antarctic Plateau, and found that 
the weather is bi-modal: 60\% of the time the sky is clear with 
low precipitable water vapor (PWV) and
weak katabatic winds (3 to $8 \, \mathrm{m\, s^{-1}}$) emanating 
from the East Antarctic Plateau; 30\% of the time is 
cloudy with higher PWV and
stronger winds (6 to $10 \, \mathrm{m\, s^{-1}}$) emanating 
from the Weddell Sea.
Incidentally, the highest wind speed at the South 
Pole measured during continuous monitoring between
1957 and 1983 was only $24~{\rm m~s^{-1}}$, and for many months
the wind speed did not exceed $12~{\rm m~s^{-1}}$.  
This low ``maximum wind speed" is favorable from a telescope 
construction standpoint.

In addition to the meteorological measures, 
the millimeter and sub-millimeter opacity at the Pole has been measured
by several experiments in the past decade
\cite{chamberlin94,chamberlin95,chamberlin97,lane98,stark01,peterson03}.  
The results show that deep millimeter-wave observations are possible most of the
time;
the median winter PWV\cite{chamberlin01} is only 0.25 mm as the
air is dessicated by frigid temperatures
(annual average: $-49~ \rm C$, minimum temperature: $-82~ \rm C$).

As mentioned previously, atmospheric stability is of extreme
importance for our measurements.  
Spatial fluctuations in the brightness of the sky result in temporal 
noise in detector timestream data as the telescope moves the beam across the sky, or as 
the wind blows the atmosphere through the beam.
As a function of angular scale, the power in atmospheric fluctuations
is well described by a Kolmogorov spectrum and 
therefore falls rapidly with decreasing angular scale.

Lay and Halverson\cite{lay98} used observations made with the Python experiment
operating at $40\,$GHz to characterize atmospheric fluctuations at 
the South Pole during
the Austral summer. 
They compared the results of the Python experiment with the site testing 
interferometer at Chajnantor\cite{radford96,holdaway95} through fits to a 
parametric model and found that the amplitude of the sky noise at the South Pole in 
the summer is 10 to 50 times less than that at Chajnantor. 
More recently, the sky noise during the Austral winter at the South Pole 
was 
characterized in detail at 
frequencies of $150$, $219$, and $274\,$GHz by the 
ACBAR\cite{runyan03a} experiment. 
Bussman {\it et al} used the correlation between ACBAR detectors to 
characterize the atmospheric fluctuation power in the presence of detector noise
over the entire winter\cite{bussman04}. 
Using the water vapor opacity predicted by 
the modified ATM code\cite{serabyn02},
we compared the ACBAR results with those of the Python experiment and
found that the median fluctuation power during Austral winter 
is approximately 20 times smaller than in the summer. 

As discussed in Section~\ref{sec:atmo}, we have used
atmospheric noise simulations normalized by 
the ACBAR measurements of atmospheric fluctuation power 
at the South Pole
to simulate SPT scan strategies in the presence of such
atmospheric noise.
The combination of the observed atmospheric power, wind 
speed, and angular scale
of interest was used to compute a 
minimum telescope scan speed below which the  
atmospheric noise becomes comparable to the detector noise.
The excellent atmospheric noise properties at the Pole enable quite
low scanning speeds (as low as $2'$/s) with only a small impact on 
our science goals;  this has in turn allowed the very clean
SPT optical design which relies on telescope scanning
(rather than a moving mirror) to move the array response across the sky.

\section{Telescope}
\label{sec:telescope}

\begin{figure}
\begin{center}
\begin{tabular}{c}
\includegraphics[height=6.85cm]{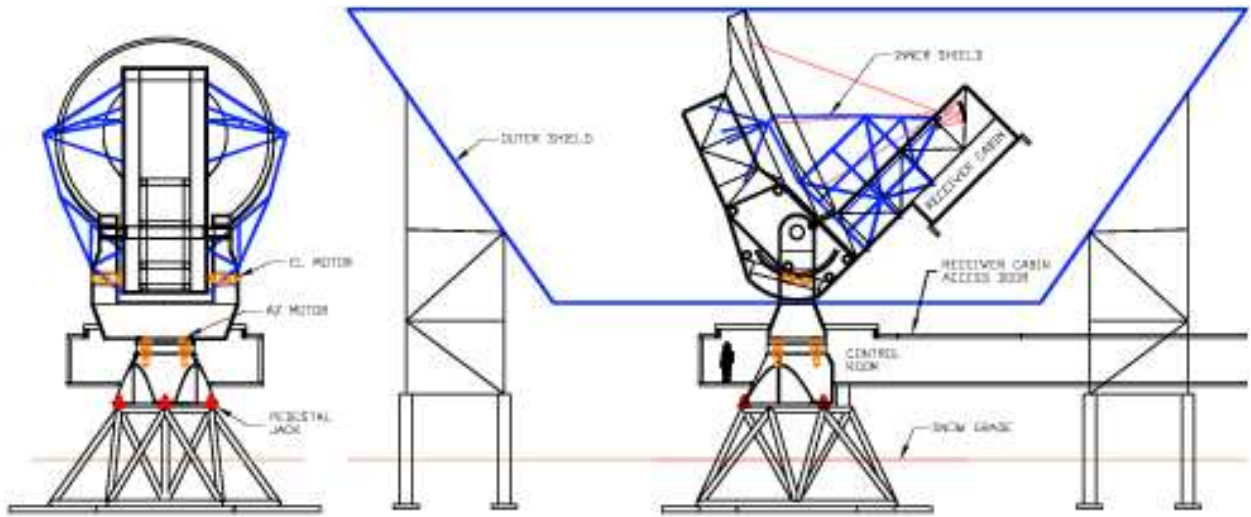}
\end{tabular}
\end{center}
\caption[sample]{
\label{fig:sptga}
(Left) rear view of the SPT at elevation $0^{\circ}$, and (right) side view at
elevation $45^{\circ}$ with the outer ground shield.}
\end{figure}

The SPT will be a 10-m off-axis Gregorian telescope on an alt-az mount. The
general arrangement is shown in Figure~\ref{fig:sptga}. The telescope
design is driven by the demanding science goals (see
Section~\ref{sec:science}) which require high sensitivity measurements
of low contrast differential emission over a broad survey region. A
large instantaneous field of view, low system noise, and stringent
control of systematic offsets such as differential pickup of thermal
ground emission are essential.

The telescope has key features designed to meet these needs:

\begin{enumerate}
\item High throughput. The SPT has a $\sim1$ deg$^{2}$ diffraction-limited
field of view at $\lambda=2$~mm, so it can support cameras with several
thousand detectors.

\item Low noise. The off-axis design gives low scattering, the gaps between
primary panels are sealed, and the beam is well shielded. There are shields
around the beam along the secondary support structure, and the entire
telescope sits inside a large, stationary, conical ground shield. The primary
is equipped with de-icing heaters to prevent ice from accumulating on the panels.  All these lead to low optical loading on the detectors, in turn leading
to high sensitivity.

\item Low offsets. The entire telescope can be chopped and scanned, so the
beam does not move on the telescope mirrors. The drive supports $2^{\circ}%
$s$^{-1}$ slew rate in elevation, $4^{\circ}$s$^{-1}$ in azimuth, $4^{\circ}$s$^{-2}$
acceleration in both axes, and position switching over $1^{\circ}$ in 1.5 s
(settling within 3 arcseconds of the required position).  In addition,
the extensive ground shielding leads to low sidelobe response in the
direction of local features (eg buildings, and the horizon).

\item Submillimeter operation. The primary has a surface accuracy of $20$
$\mu$m rms, and the pointing accuracy of the telescope is $1.5$ arcseconds
rms, so operation at wavelengths as short as $200$ $\mu$m will be possible.
\end{enumerate}

The telescope is being built by VertexRSI. It will be assembled and tested in the
US in Jan--May 2006, prior to deployment at the South Pole in Nov 2006--Jan 2007.

\subsection{Optical Design}
\label{sec:optics}

We have chosen a classical Gregorian design for the SPT to provide flexibility
for future optical configurations.  In particular, this design can
accommodate a chopping mirror at the pupil after the secondary, and in\ the
classical Gregorian design (with a paraboloidal primary) the focal ratio of the
telescope can be changed simply by replacing the secondary.  The off-axis form
was chosen because it allows a large secondary, and hence a high throughput,
with low scattering and high efficiency.

Millimeter-wave Gregorian telescope designs typically have a chopper at the pupil just
after the secondary.  This arrangement requires re-imaging optics for low
aberations over a wide-field of view\cite{stark00,stark03b}.  Unfortunately,
the many reflections in these designs lead to degraded
sensitivity ($\sim1\%$ loss per warm mirror, causing extra loading on the
detectors), high instrumental polarization, and chop-synchronous offsets (due
to the beam moving on the secondary). Since the SPT is capable of fast
scanning, we have adopted a much simpler optics configuration in which the
camera is located at the Gregorian focus and there is no chopping mirror. In
this compact two-mirror scheme, the stop is at the secondary (as in a typical
infrared telescope) and spillover on the stop is absorbed by a cold load which
surrounds the beam between the camera and the secondary.

\begin{figure}
\begin{center}
\begin{tabular}{c}
\includegraphics[height=7.15cm]{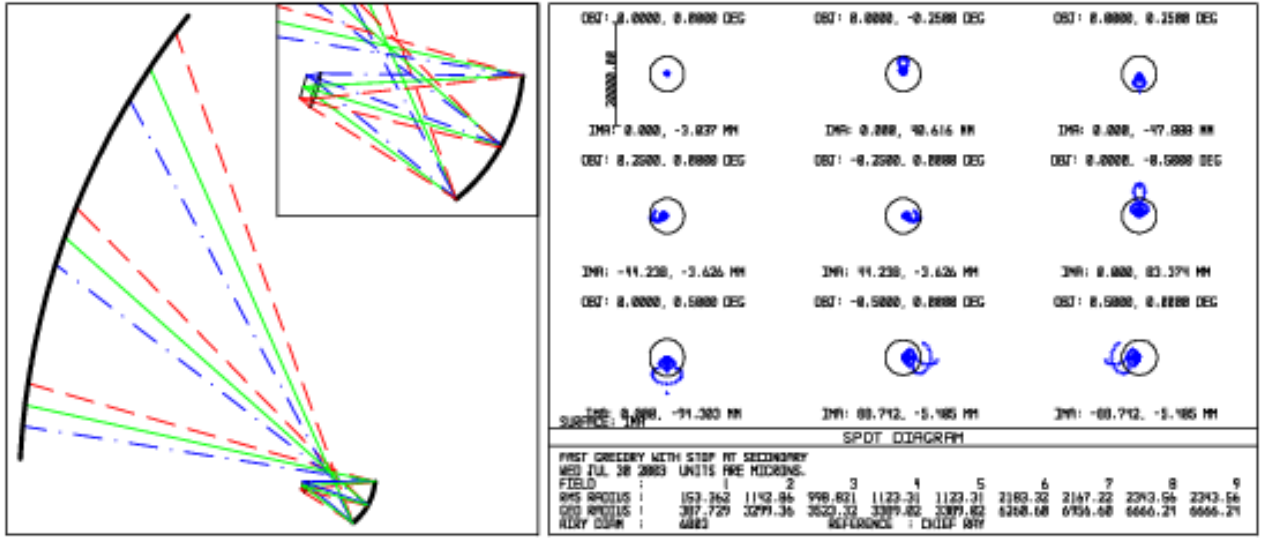}
\end{tabular}
\end{center}
\caption[sample]{
\label{fig:sptoptics} 
(Left) optical configuration of the SPT. Solid lines are the principal and
marginal rays for the field center. Dashed and dot-dashed lines are for the
edges of a $1^{\circ}$ diameter field. (Right) spot diagrams. Circles show
the Airy disc at $\lambda=2$ mm.}
\end{figure}

The optical configuration of the SPT, and corresponding spot diagrams, are
shown in Figure~\ref{fig:sptoptics}. The Gregorian focus is very fast --
$f/1.3$, which gives maximum coupling to feedhorns of diameter $2f\lambda=5$
mm at $\lambda=2$ mm \cite{griffin02}. In this case, a 1000-element bolometer
array is $\sim150$ mm in diameter. The secondary is only 1 m in diameter,
which is important because: (i) it sets the size of the cryostat for the cold
stop; and (ii) the difficulty and cost of making a monolithic secondary
increases rapidly with diameter if the secondary is larger than $\sim1$ m. The
Gregorian focus is stigmatic at the field center, and the geometric cross polarization is
zero, i.e., the system satisfies the Dragone condition \cite{dragone82}. At a
field radius of $0.5^{\circ}$, the peak cross-polar response is roughly $-30$ dB.

The secondary and the entire beam path from prime focus to the camera is
cooled, giving low noise and stable spillover. The loss of the telescope is
also low because there is just one warm reflection (at the primary) and one
warm transmission (through the cryostat window near prime focus). The penalty
for placing the stop at the secondary is $\sim10\%$ degradation in resolution
because the illumination pattern on the primary varies with field position. The
entrance pupil of the SPT is $56$ m in front of the primary, so the
illumination pattern at the edge of a $1^{\circ}$ diameter field is displaced
$\sim0.5$ m. Thus, the diameter of the illumination pattern must be $\sim1$ m
smaller than if the entrance pupil were at the primary.

\begin{figure}
\begin{center}
\begin{tabular}{c}
\includegraphics[height=7.15cm, bb=0 250 560 555]{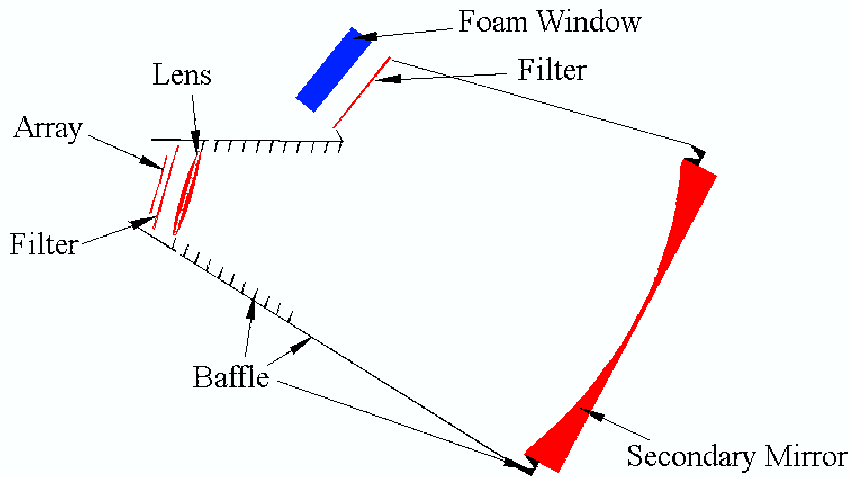}
\end{tabular}
\end{center}
\caption[sample]{
\label{fig:baffle} 
A cutaway view of the cold secondary and 10~K baffle optical 
configuration.  Radiation from the primary 
enters through a foam vacuum window from the upper left,
passing through an IR blocking filter that reduces 
radiative heat input on the 10~K system.  The
baffle is formed by two metal cones coated with a millimeter-wave
absorber, with black annular rings attached to the cone wall near 
the 4~K lens and focal plane.  For a Gaussian beam with 6~dB edge taper
and a baffle emissivity of 0.5, ray-tracing of this design shows that less 
than 1\% of the spillover power eventually exits the cryostat window.
}
\end{figure}

Roughly $25\%$ of the power from a $2f\lambda$ diameter, smooth-wall, conical feedhorn
spills over the secondary and 
falls on the cold stop, so emission from the stop can cause significant
loading. The temperature of the cold stop in the SPT is $\sim10$ K, which
gives a reasonable compromise between sensitivity loss ($\sim5\%$ at the
zenith at $\lambda=2$ mm) and the difficulty of cooling a large mirror
assembly.  As shown in Figure~\ref{fig:baffle}, the cold stop is 
a conical absorbing shroud, with annular baffles
near the Gregorian focus. The secondary is a lightweighted mirror
(total mass $\sim15$ kg), machined from an aluminum plate, with thermal
cycling between machining steps to reduce deformation of the finished mirror
on cooling. The secondary and stop assembly is supported by a truss attached
to the cryostat wall near the prime focus port. Since the SPT must support
cameras for different wavelengths, and also a polarimeter, the cryostat is
split into two parts, one containing the secondary and cold stop, the other
containing the detector array. The two cryostats share the same vacuum, but
have independent refrigerators.

A lens immediately in front of the detector array makes the focus telecentric
to improve coupling to the feedhorns. This lens is quite weak ($\sim f/10$),
so it can be made from a material with fairly low refractive index, e.g.,
high-density polyethylene. A silicon lens would give lower loss, but a
suitable anti-reflection (AR) coating is not yet available. A wide-band AR
coating for the plastic lens is also not entirely straightforward. We are
considering machined, profiled grooves and a multi-layer stepped-index coating
made of thin sheets of laser-drilled plastic.

A key advantage of the SPT observing strategy is that the entire telescope is
scanned, so the beam does not move relative to the telescope mirrors. This
approach reduces offsets, but in practice there will be some scan-induced
deflections and associated changes in spillover. The stability of the beam on
the window and heat blocking filter near prime focus is a particular concern
because the beam there has fairly sharp edges.  For a filter over-sized by
10~mm in radius, a displacement of the filter
changes the power at the detector by $\sim30$ mK mm$^{-1}$.  The receiver
noise is $\sim100$ $\mu$K Hz$^{-1/2}$, so scan-asynchronous changes 
of more than a few micrometers in the position of the beam on 
timescales corresponding
to the signal band will degrade the sensitivity.  
Temperature fluctuations in the cold stop also
add noise, but typically will not be synchronous with the telescope scan. 
Again to avoid degrading sensitivity, such fluctuations much be below
the level of a few
$\times10$ $\mu$K Hz$^{-1/2}$ averaged over the detector readout bandwidth.

\subsection{Primary}
\label{sec:primary}

The 10-m f/0.7 SPT primary design has lightweighted machined aluminum panels (each
roughly $0.75\times0.5$ m), mounted on a carbon fiber reinforced plastic
(CFRP) back-up structure (BUS). The panels will have $\sim8$ $\mu$m rms surface
accuracy.  Differential contraction between the BUS\ and panels makes the panel
gaps wider at lower ambient temperatures; if unaddressed, the effect of these
gaps (emission and scattering)
would increase as
the atmosphere improves.  At the South Pole, the temperature of the primary
could be as high as $-10^{\circ}$C in summer, and to prevent the panels from
touching at this temperature the panel gaps must be $\sim2$ mm at $-80^{\circ
}$C. Emission from the gaps, and offsets due to small movements of the beam
relative to the gaps, are a serious concern, so we plan to cover the gaps with
$\sim0.1$ mm thick metal strips held in place by spring fingers.

Ice on the primary panels increases the telescope emission, and causes
slowly-varying offsets, and ice in the panel gaps can change the alignment of
the panels. Some telescopes at the South Pole have employed manual
de-icing techniques, but these are impractical for a 10-m telescope. The SPT
primary will be equipped with de-icing heaters, e.g., heat blankets on the
BUS, or hot air blowing between the BUS and the panels. In addition, we will
select a panel surface finish that discourages ice accumulation. We are
currently testing a variety of finishes at the South Pole, e.g., polished, etched, anodized, and
SiO$_{2}$ coated. Initial results from this test indicate
that $\sim50$ Wm$^{-2}$K$^{-1}$ is required to heat the panels, a few K
temperature rise is enough to de-ice the panels on timescales of a day or two,
and the different surface finishes have roughly similar icing characteristics
(but the coated surfaces are generally colder than the bare aluminum
surfaces). Since de-icing can be done with a very small temperature increase,
we are currently planning to run the de-icing heaters continuously at low
power. This avoids the problem of having to wait for the primary to recover
its profile after high-power de-icing, and reduces the peak demand on the
South Pole power plant.

The SPT\ primary has 217 panels, each with 4 axial and 3 in-plane
adjusters. The surface will be aligned initially to $\sim60$ $\mu$m rms based
on photogrammetry measurements, and later to $20$ $\mu$m rms based on
holography. Setting all the panel adjusters under the conditions at the South
Pole will be a difficult task, so access to the adjusters is from inside the
BUS. This is a cramped environment, but it provides protection from the wind,
and allows access to the adjusters without a crane or man lift.

\subsection{Ground Shields}
\label{sec:shields}

Low noise is a critical requirement for the SPT, so the design includes
several levels of shielding. The inner shield is a trough running along the
secondary support as shown in Figure~\ref{fig:sptga}. Gaps between the inner shield,
primary and secondary support are sealed with reflecting material, and the top
edges of the shield and primary are rolled with a radius of a few cm to reduce
scattering. The inner shield does not have de-icing heaters, so it will
require occasional brushing to remove snow (through doors at the bottom of the
shield near the primary).

The outer shield is an enormous inverted cone around the entire telescope. The
wall angle of the cone is shallow enough to prevent radiation being trapped
between the shield and the (reflective) back of the primary, so the shield
does not require a floor. This makes snow removal much easier. The shield is
made of 1.25-mm thick aluminum panels $\sim3$ m across (the width of an LC130
aircraft) supported by a steel spaceframe. The panels are finished with a
smooth epoxy phenolic coating to encourage snow shedding. A $60^{\circ}$ wide
section of the shield can be lowered to allow short, occasional observations
near the horizon, e.g., calibration observations of planets, and holography
using a test source on a tower. Ray tracing of the telescope with inner and
outer shields shows that variations in ground pickup will be $<1$mK per
degree change in elevation.

\subsection{Mechanical design}
\label{sec:mechanical}

The SPT primary and secondary are mounted on a massive L-shaped frame on an
alt-az fork mount (see Figure~\ref{fig:sptga}). The mount is made entirely of steel,
and the CRFP BUS is attached to the steel L-frame via an Invar cone. The mount
is balanced about both axes to minimize deflections in the structure and
settling of the foundation, but this requires a large elevation counterweight.
The azimuth bearing is at the top of a conical pedestal, which is supported by a
massive spaceframe sitting on wood footings. The base of the pedestal is
$\sim4$ m above the ice to reduce snow drifting around the structure.
Table~\ref{tab:mass}
shows the masses of the major components of the SPT. Each component is
designed so that it can be broken down into parts that will fit in an
LC130\ aircraft (which can carry about 11 metric tons).

Each axis has two pairs of torque-biased motors. The azimuth motors drive a ring
gear inside the azimuth bearing, and the elevation\ motors drive sector gears on each side
of the L-arm. The yoke and L-frame contain CFRP reference structures with
displacement sensors to measure gravitational and thermal deflections of the
mount. An active optical bench, which is coupled to the reference frame
system, supports the secondary and camera. In this configuration, the relative
positions of the secondary, lens and detector array are fixed and the entire
assembly is continuously adjusted (with a control bandwidth of $\sim0.1$ Hz)
to maintain its position with respect to the primary. The mount also has
tiltmeters above and below the azimuth bearing to measure changes in the tilt of
the structure (which are included in the pointing model with a bandwidth of
$\sim0.1$ Hz). All the SPT\ drive components are modular, moving parts and
critical electronics are in warm environments, and the drive motors,
gearboxes, encoders, reference frame sensors and tiltmeters can be reached
from the control room without going outside. The outside of the mount is
insulated to reduce differential cooling, and interior spaces in the pedestal
and fork are warmed by air from the control room.

The receiver cabin is a large ($\sim6$m long $\times$ 3m high $\times$ 2.5m wide)
shielded and
insulated room at the end of the L-frame. This provides a laboratory
environment for optics and cameras, with enough space to accommodate future
optical configurations with many mirrors and several optical benches. A door
in the floor of the cabin mates with a hatch in the control room roof when the
telescope is at the horizon (see Figure~\ref{fig:sptga}). Access to the
cabin is also possible from outside, through the L-frame, in any telescope
position.

\begin{table}
\caption{\label{tab:mass}Expected masses of SPT components.}
\begin{center}
\begin{tabular}
[c]{lc}%
Component & Mass (metric tons)\\ \hline
Foundation frame + footing & 34\\
Pedestal & 23\\
Fork & 16\\
L-frame & 77\\
Counterweight (lead) & 73\\
Primary (Invar cone + BUS +panels) & 18\\
Secondary + camera & 3\\ \hline
Total & 244
\end{tabular}
\end{center}
\end{table}

\section{Receiver}
\label{sec:receiver}

The statistical arrival of photons presents a fundamental limitation to the 
sensitivity of bolometric detectors, which has
been reached for small ground-based arrays of detectors 
operating at mm-wavelengths\cite{runyan03a}. 
The SPT science objectives require significant advances
in mapping speed; to achieve this, the SPT receiver will be based on
a focal plane with 1000 superconducting Transition Edge Sensor (TES) bolometers.
Observations will be done sequentially in at least three frequency bands to 
enable spectral separation of various celestial sources.
Reading out 1000 cryogenic detectors 
poses a technological challenge;
we plan to use a SQUID-based frequency domain multiplexer
readout to minimize the complexity and heat load of cold wiring.
To minimize maintenance and cost of operation, the bolometer camera 
will be cooled by the combination of a $250\,{\rm mK}$ closed cycle 
refrigerator and a 
mechanical refrigerator that reaches $2.3\,$K without liquid cryogens.  
In the following sections we describe the detector array, cryogenics, readout
technology and cryostat optics.

\subsection{Focal Plane Array}
\label{sec:array}

The SPT focal-plane array will use horn-coupled spider-web bolometers with superconducting
Transition-Edge Sensors (TES).  
Compared with conventional semiconductor-based technologies, TES detectors offer
several advantages for the construction of large arrays.
First, the bolometers are produced entirely by thin-film deposition and optical 
lithography, greatly simplifying fabrication and improving the uniformity of devices.  
Second, readout multiplexing technologies have been developed 
for TES detectors that have the potential  
to greatly reduce the complexity and cost of large arrays.
Other benefits include high linearity and well controlled 
responsivity
independent of base temperature and optical-loading, 
due to the TES's strong electrothermal feedback effect.
In addition, 
voltage biased TES bolometers are also quite 
insensitive to vibration by virtue of their low impedance.

\begin{figure}[bt]
\begin{minipage}{.475\textwidth}
\centerline{\includegraphics[height=2in,bb=0 0 570 430]{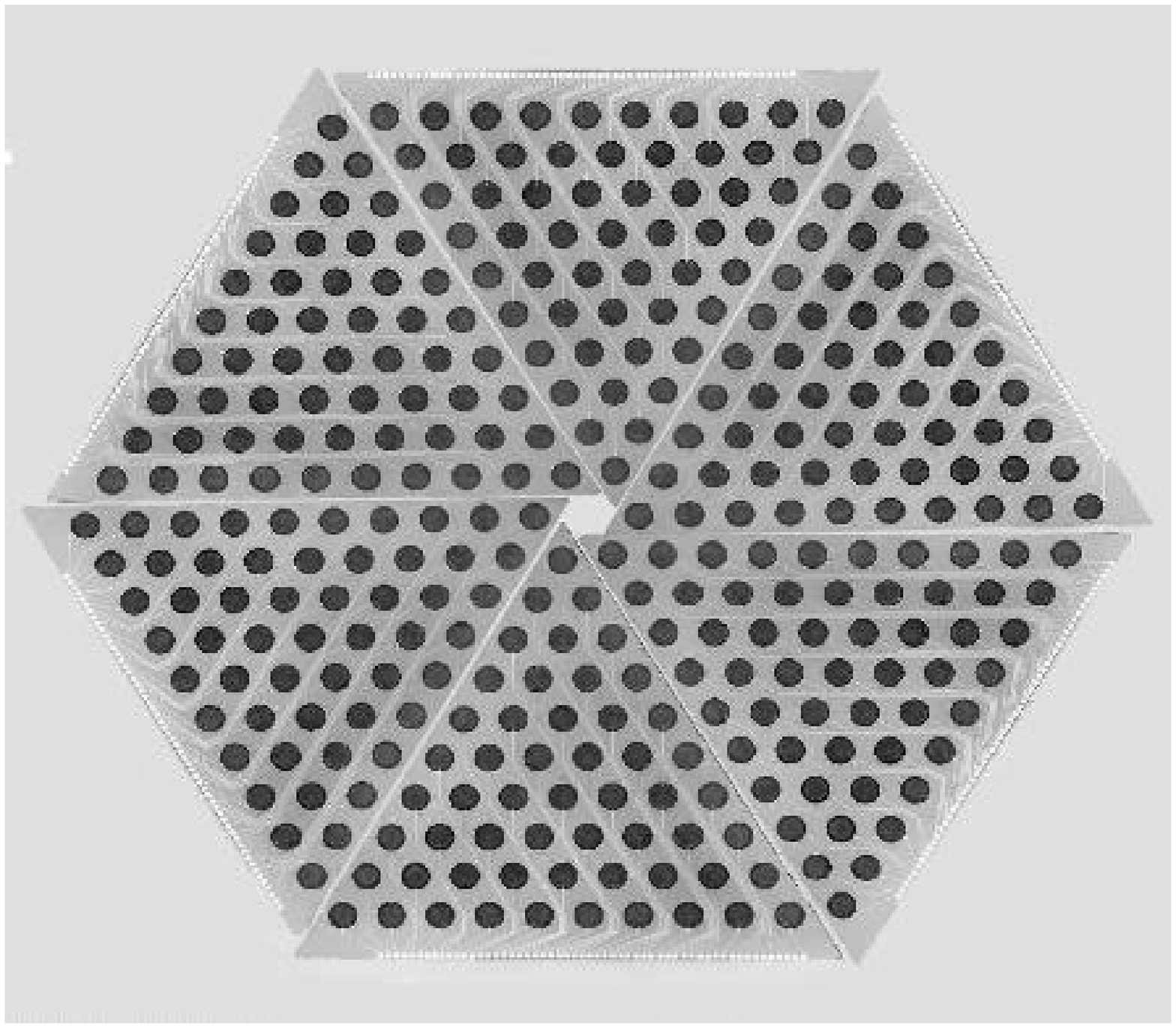}}
\caption[]{ Montage image of a single 55-element TES spider
bolometer wedge to show how an  array of six identical wedges would look.  
The complete prototype array will have 330 bolometers and be 12~cm
in diameter. } 
\label{fig:array}
\end{minipage}
\hfill
\begin{minipage}{.475\textwidth}
\centerline{\includegraphics[height=3.2in,angle=-90]{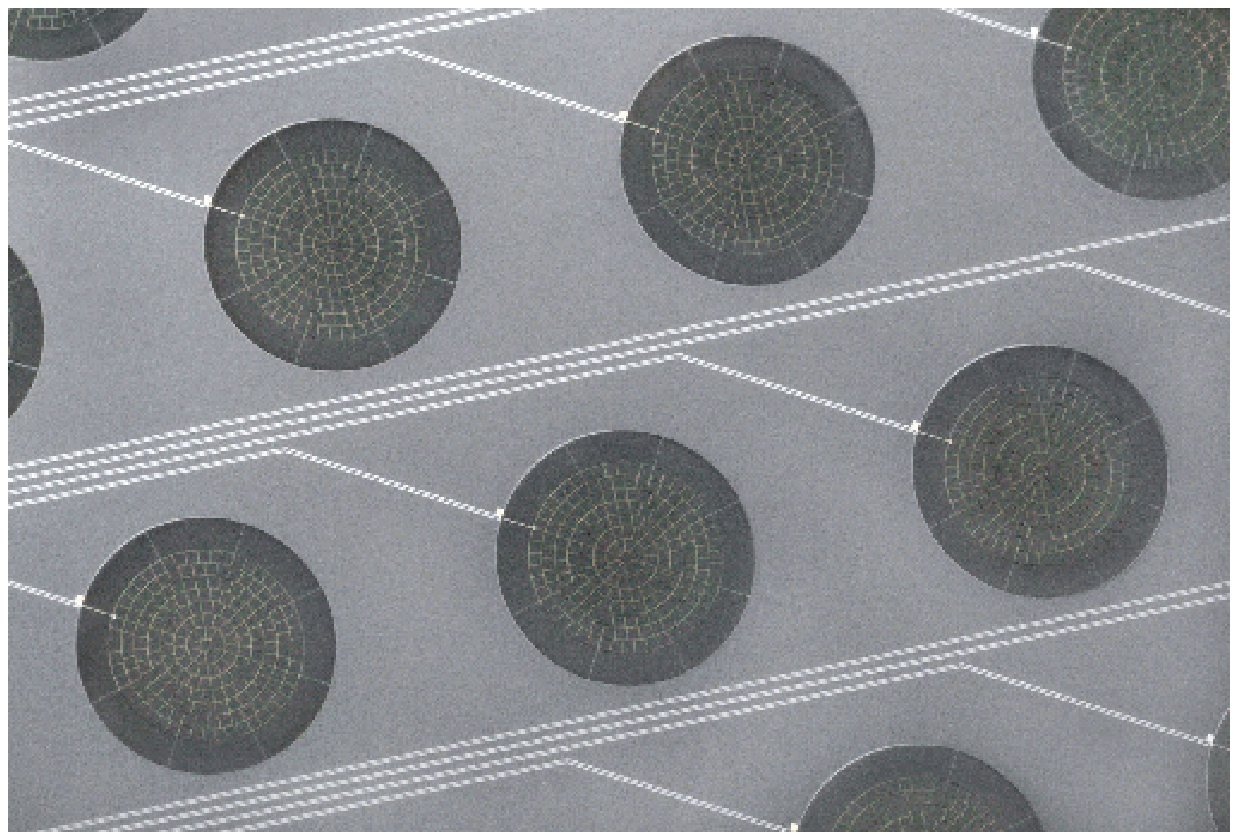}}
\caption[]{ Close up of
a 55 element bolometer wedge.  
The sensors are constructed with an Al/Ti proximity effect 
sandwich. Webs are metalized with gold for microwave
absorption.  Suspended spider-web absorbers
are fabricated from 1 $\rm{\mu m}$ thick silicon nitride.  The
membrane is released from the front side using a gaseous xenon
diflouride etch.  Bolometers are 5 mm diameter with 0.5 mm long legs.
Wiring layer is superconducting aluminum. This array was fabricated in
the U.C. Berkeley microfabrication facility.} 
\label{fig:closewedge}
\end{minipage}
\end{figure}

The SPT array will be assembled from 6 pie-shaped wedges
each with approximately 160 bolometric detectors. 
Each detector consists of a
small Al/Ti bi-layer TES suspended on a gold-covered Silicon Nitride spider-web absorber. 
We have fabricated 55 element prototype wedges of TES
detectors on 4 inch wafers with high yield, 
like those shown in Figures~\ref{fig:array} and \ref{fig:closewedge}.
The individual detectors in the 55 element wedges meet the requirements for 
the SPT array, though
the 160 element SPT wedges will need to be built on 6 inch diameter 
wafers.

The sensor is a bi-layer of Aluminum and Titanium, with layer thicknesses
tailored to give a transition temperature of $0.5\,$K. 
In order for the detectors to operate with strong electrothermal feedback,
the thermal conductivity needs to be chosen such that 
when they are subjected to the maximum expected optical load, the
electrical power applied to keep the sensor in the transition region
is comparable to the optical power.   
Unfortunately the optical background can 
be hard to predict and depends on the frequency band and elevation of the 
observations. 
From experience with ACBAR, we estimate the emission from cryostat optics, 
filters, vacuum window, and the warm primary mirror 
(taken to be 1\% emissive) will be 
approximately $20\,$K in all frequency bands.
We conservatively set the thermal conductivities $G$  to values 
appropriate for 3-5 times more electrical power than the expected total optical 
loading during observations at elevation $60^\circ$.
Even with this conservative design, the photon noise completely dominates
the total noise of the detectors.
Compared with NTD bolometers, the TES detectors have a much smaller 
contribution from Johnson noise in the sensor, and therefore somewhat lower 
total noise. 
Table~\ref{tab:sensitivity} summarizes the design specifications
and expected performance for the bolometers in each of the 
potential observation bands.
In the $2.1\,$mm band, the array will image one square degree of sky
to an RMS of $\Delta T_{\rm cmb} \sim 10\,\mu$K in a single hour of
observation.

\begin{table}
\begin{center}
\vspace{0.2in}
\begin{tabular}
[c]{ccccccccc}
$\nu_0$ & $\Delta \nu$ & T & $P_o$ & G  & NET$_{\rm RJ}$ & NET$_{\rm CMB}$ & $\theta_{fwhm}$ & NEFD  \\ 
(GHz) & (GHz) &   & (pW) & (pW/K) & ($\mu{\rm K} {\sqrt {\rm s}}$) & ($\mu{\rm K} {\sqrt {\rm s}}$) & (arcmin) & (mJy $\sqrt{\rm s}$) \\ \hline 
95  & 24 & 0.964 &  8.1 & $2\times 10^{-10}$ & 221 &  278  & 1.58 & 14.6 \\ 
150 & 38 & 0.982 & 10.8 & $2\times 10^{-10}$ & 150 &  259  & 1.00 & 9.9 \\
219 & 35 & 0.969 & 11.0 & $2\times 10^{-10}$ & 184 &  551  & 0.69 & 12.2 \\
274 & 67 & 0.950 & 24.5 & $4\times 10^{-10}$ & 159 &  774  & 0.56 & 10.5 \\
345 & 27 & 0.844 & 22.3 & $4\times 10^{-10}$ & 425 & 4975  & 0.44 & 28.1 \\
\end{tabular}
\end{center}
\caption{Detector specifications and expected performance in each of the potential 
observations bands; the current baseline for the SPT is to observe in the 
150, 219 and 274 GHz bands, subject to ongoing simulations of foreground
removal.
$\nu_0$ is the center of the frequency band and $\Delta \nu$ is the bandwidth. 
T is average transmission of the atmosphere in the band. 
$P_o$ is the total optical power in the band for an observation 
at elevation $60^{\circ}$ and an optical efficiency of 40\%. 
G is the thermal conductivity of the detector. 
NET$_{\rm RJ}$ and NET$_{\rm CMB}$ are the noise equivalent 
temperatures in Rayleigh-Jeans and CMB temperature units. $\theta_{fwhm}$ is the 
full width at half-maximum of the diffraction limited telescope beam and NEFD 
is the noise equivalent flux density.
\label{tab:sensitivity}
}
\end{table}

\subsection{Cryogenics}
\label{sec:cryogenics}

The baseline for the SPT receiver is to use a system 
free of expendable cryogens, which has strong benefits for
operations at the South Pole.
We have assembled a
test system using a Cryomech model PT-405 pulse tube cooler which 
provides $0.7\,$W of
cooling power at $4.2\,$K and a base temperature of 2.3K.  
A Chase Research 3-stage Helium
sorption refrigerator operates from the pulse tube cold stage.  It 
has one stage of $^4$He used to condense $^3$He in two separate
reservoirs, the warmer of which acts as a buffer for the 
coldest stage.
The entire sorption refrigerator cycling process takes 
about 2 hours; the cold stage then
maintains a steady temperature of $250\,$mK for 56 hours under an
external heat load of $1\,\mu$W.
The buffer $^3$He stage operates at
$350\,$mK with a cooling power of $100\,\mu$W; this stage is used to
intercept heat loads from wiring and mechanical supports. 
We have found that vibrations from the pulse-tube cooler do not excite 
a substantial microphonic response in the TES bolometers, and we
can achieve the baseline detector noise with the pulse-tube operating.

We have also made detailed measurements of the temperature
fluctuations on each of the cryogenic stages.  The fluctuations at
$250\,$mK and $50\,$K are small enough to be negligible in the total 
detector noise budget.  However, at $3\,$K the fluctuations 
from the cycling of the pulse tube can be as large as $50\,$mK.
It will be necessary to control the temperature of the $3\,$K lens
and IR blocking filter to a level of approximately 
$1\,$mK so that this signal is below the noise floor of the detectors. 
This will be achieved with a
passive thermal circuit that makes use of a weak link to the large heat capacity
of the cold lens. 
A gas gap heat switch will be used to speed the initial cooling of the 
lens and filter assembly from $300\,$K. 

\subsection{Receiver Optics}
\label{sec:rec_optics}

\begin{figure}[hbt]
\begin{minipage}{.475\textwidth}
\centerline{\includegraphics[height=2.0in]{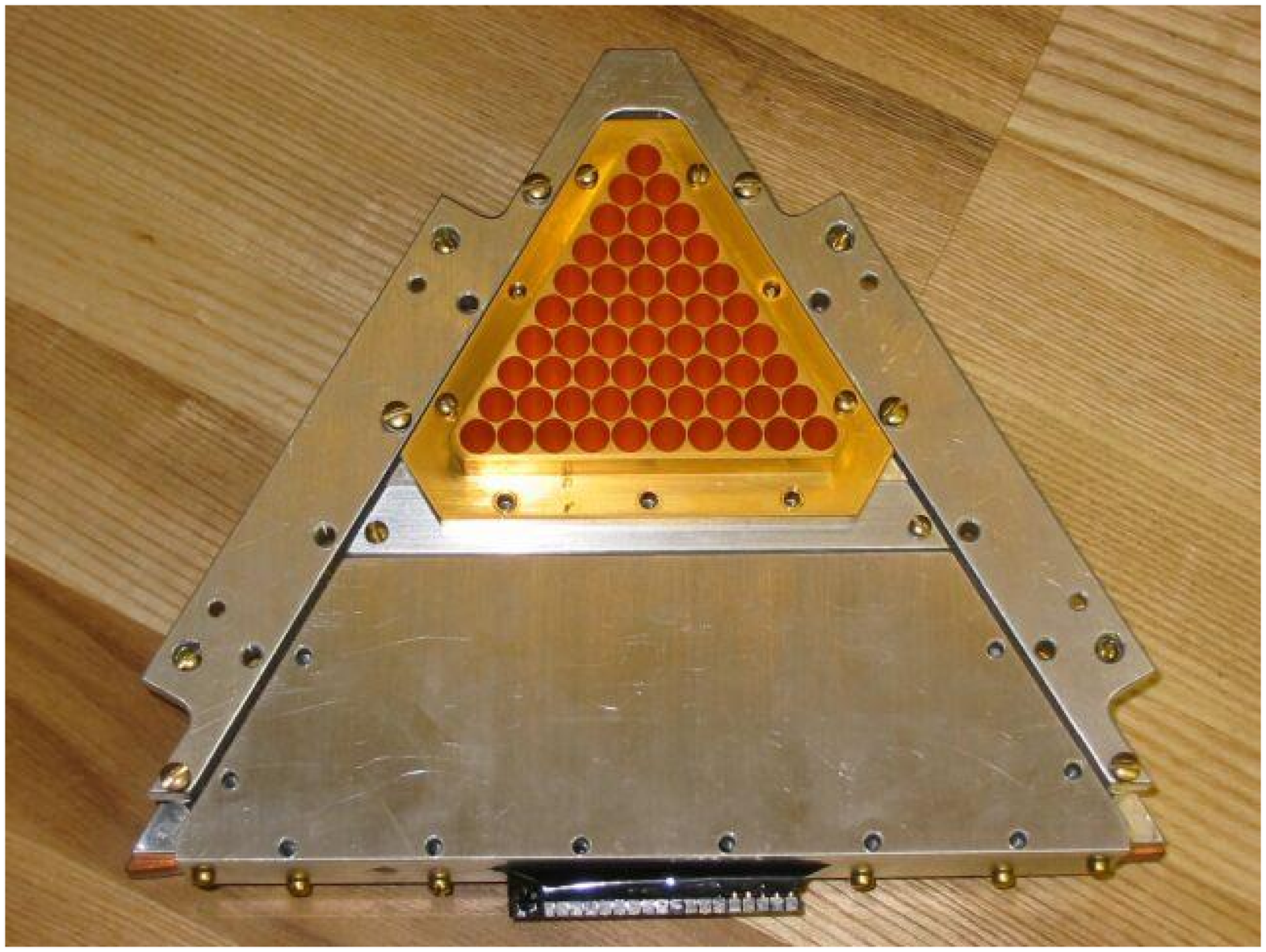}}
\caption[]{ A 55 element horn array mounted on top of a prototype 
detector wedge.  The SPT array will have approximately 160 horns and
detectors per wedge.} 
\label{fig:hornarray}
\end{minipage}
\hfill
\begin{minipage}{.475\textwidth}
\centerline{\includegraphics[height=2in]{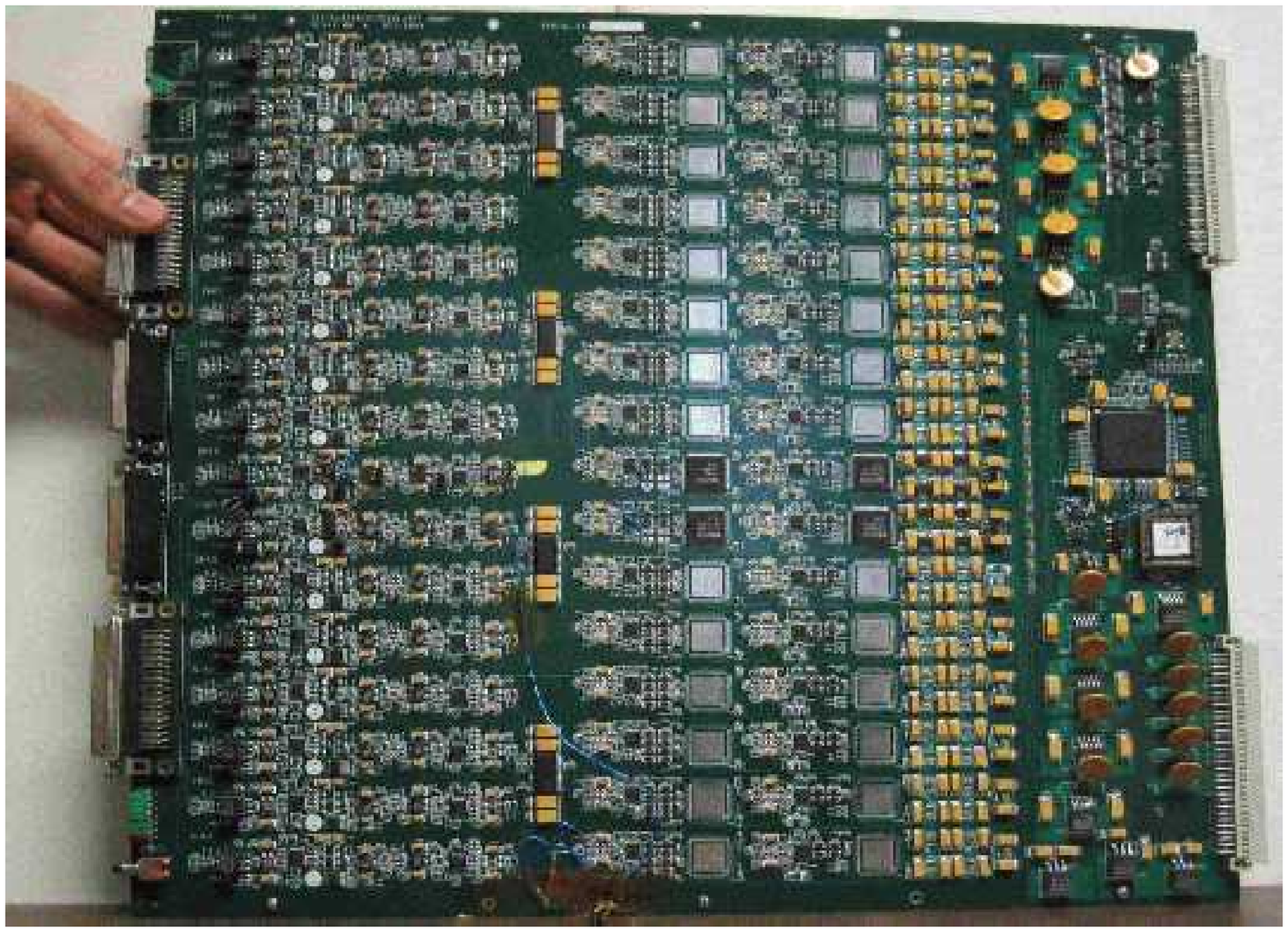}}
\caption[]{ The 16-channel oscillator/demodulator boards provide the AC bias sine-wave carriers 
which are sent to the bolometers.  The carriers with 
associated sky-signal amplitude modulation 
are received from the bolometers and SQUID amplifiers, mixed down to base band, and 
digitized on-board. 
One oscillator/demodulator channel is required per bolometer, therefore, 60 boards in 
three 9U VME crates are needed for the SPT receiver.} 
\label{fig:oscboards}
\end{minipage}
\end{figure}

Due to the wide field of view of the SPT, the receiver optical elements 
are quite large and present considerable fabrication challenges.
In Figure~\ref{fig:baffle}, we show a concept drawing
of the cold optics box and receiver. 
An array of smooth wall conical feeds is used to couple the incoming 
radiation to the detector array. 
The secondary acts as a cold stop and truncates the side lobes produced 
by the conical feeds.
Figure~\ref{fig:hornarray} shows
a prototype 55 element conical
horn array, fabricated from solid Aluminum
with a set of custom ground reamers and then gold plated. 
The technology used
to produce this prototype is easily scalable to the 160 element
wedges planned for the SPT.

A relatively low-power lens is used to convert the incoming beam to
the appropriate f-1.3 feed for the horns.
This lens will be fabricated from polyethylene with an
anti-reflection (AR) coating. 
The IR blocking and lowpass band-edge defining filters are constructed by bonding 
conductive mesh layers on polyethylene films in a heated press. 
The SPT optics require filters between $200$-$250\,$mm in diameter, 
the size of which presents a serious challenge 
but is within the capabilities of the fabrication facilities at Cardiff.
The lower edges of the frequency bands are set by the cutoff frequency
of a small section of circular waveguide behind the conical horns. 
The conical horn array, filters and possibly the AR-coated lens 
will need to be changed 
in order to change observation bands.
We are planning to make these systems as modular as possible to facilitate 
rapid band changes. 

As can be seen in Figure~\ref{fig:baffle}, the cold optics
design incorporates a relatively large vacuum window. We have
recently built and tested a $12"$ diameter window appropriate for
use with the SPT. 
The window is constructed from a $3"$ thick laminate of Zotefoam
PPA-30 nitrogen expanded polypropylene.
It appears to be mechanically robust and holds an excellent
vacuum. 
The total loss through the window has been measured to be less 
than $1.0\%$ at $150\,$GHz.

\subsection{Frequency Multiplexed SQUID Readout}
\label{sec:mux}

One key to implementing an array of more than several hundred bolometers is readout 
multiplexing, which can dramatically reduce the heat load,  the complexity of cryogenic 
wiring, and the cost.  The SPT utilizes a frequency-domain multiplexer which requires only a 
single SQUID to read out a module of several bolometers \cite{Spieler02,Lanting04}.  
The number of bolometers per readout module is still uncertain; eight has been
demonstrated and 32 appears to be practical, so each wedge of the array will
require several readout modules. 

\begin{figure}\centering
\includegraphics[width=\textwidth]{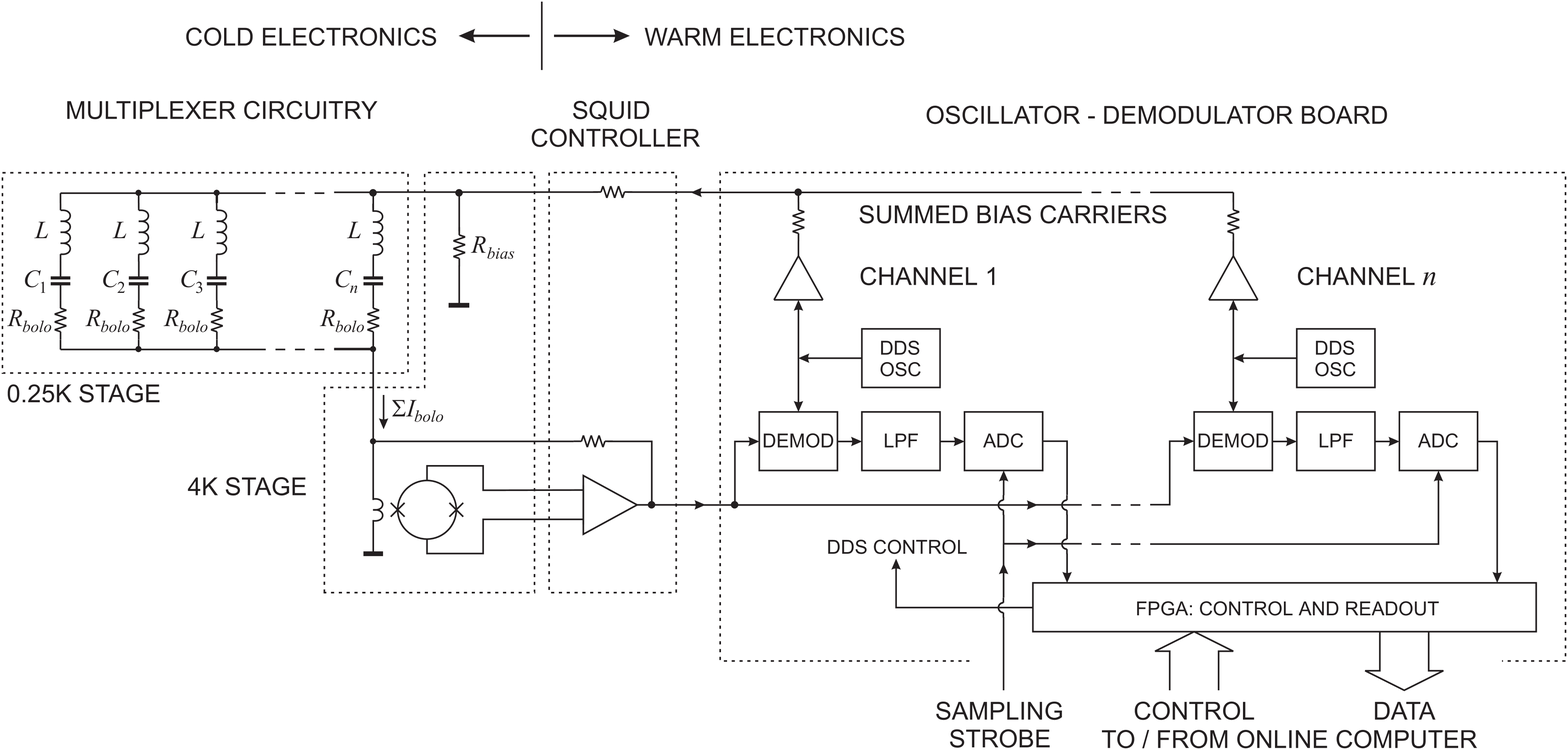}
\caption[]{ Schematic diagram showing the basic components of the
  frequency multiplexed SQUID readout system.} \label{fig:muxBlockDiagram}
\end{figure}

A schematic diagram showing the basic components of the readout system
is shown in Figure~\ref{fig:muxBlockDiagram}. 
The bolometers are sine-wave biased with a constant voltage amplitude 
carrier, in the frequency range of 500 kHz to 1 MHz. 
Each bolometer within a readout module is biased at a different frequency. 
The sky-signal changes the bolometer resistance and amplitude modulates the bolometer
current such that the signal from each bolometer is transferred to sidebands adjacent to
its carrier.
Thus, the signals from different bolometers within a module are uniquely positioned in frequency, 
so they can be summed and connected through a single wire to a SQUID amplifier. 

Each bolometer is part of a series-resonant LC circuit, which is tuned to the appropriate 
bias frequency. 
This allows the bias frequencies for all bolometers in a module to be applied through a 
single wire, as the tuned circuit selects the appropriate frequency for each bolometer. 
Thus only two wires are needed to connect the bolometers of a readout module on 
the $0.25\,{\rm K}$ stage to the $4\,{\rm K}$ stage on which the SQUIDs are mounted. 
The tuned circuits also limit the bandwidth of the bolometer's Johnson noise, which would 
otherwise contribute to the noise in all other channels of the module. 
The comb of amplitude modulated carriers at the SQUID output is transmitted to a bank of 
demodulators that mix the signals back down to base-band. 
The signals are then filtered and digitized, and all outputs 
in the array are sampled synchronously. 
Multiplexing of 8 channels has been demonstrated and the limit to the number of 
channels per readout module is being explored. The high-Q superconducting LC filters 
use lithographed inductors and commercial NP0 ceramic chip capacitors.  
All inductors have the same value, so the $Q$ increases with frequency to maintain 
constant bandwidth. 
A 16~$\mu$H inductor together with the bolometer resistance determines the bandwidth 
(5~kHz), and the capacitor sets the frequency of each channel. 
Both the inductors and chip capacitors are compact and together occupy an area 
comparable to a single pixel in the array. 
The filters will be mounted on a board underneath the bolometer array.

\begin{figure}[tb]
\begin{minipage}{.475\textwidth}
\centerline{\includegraphics[width=3.0in]{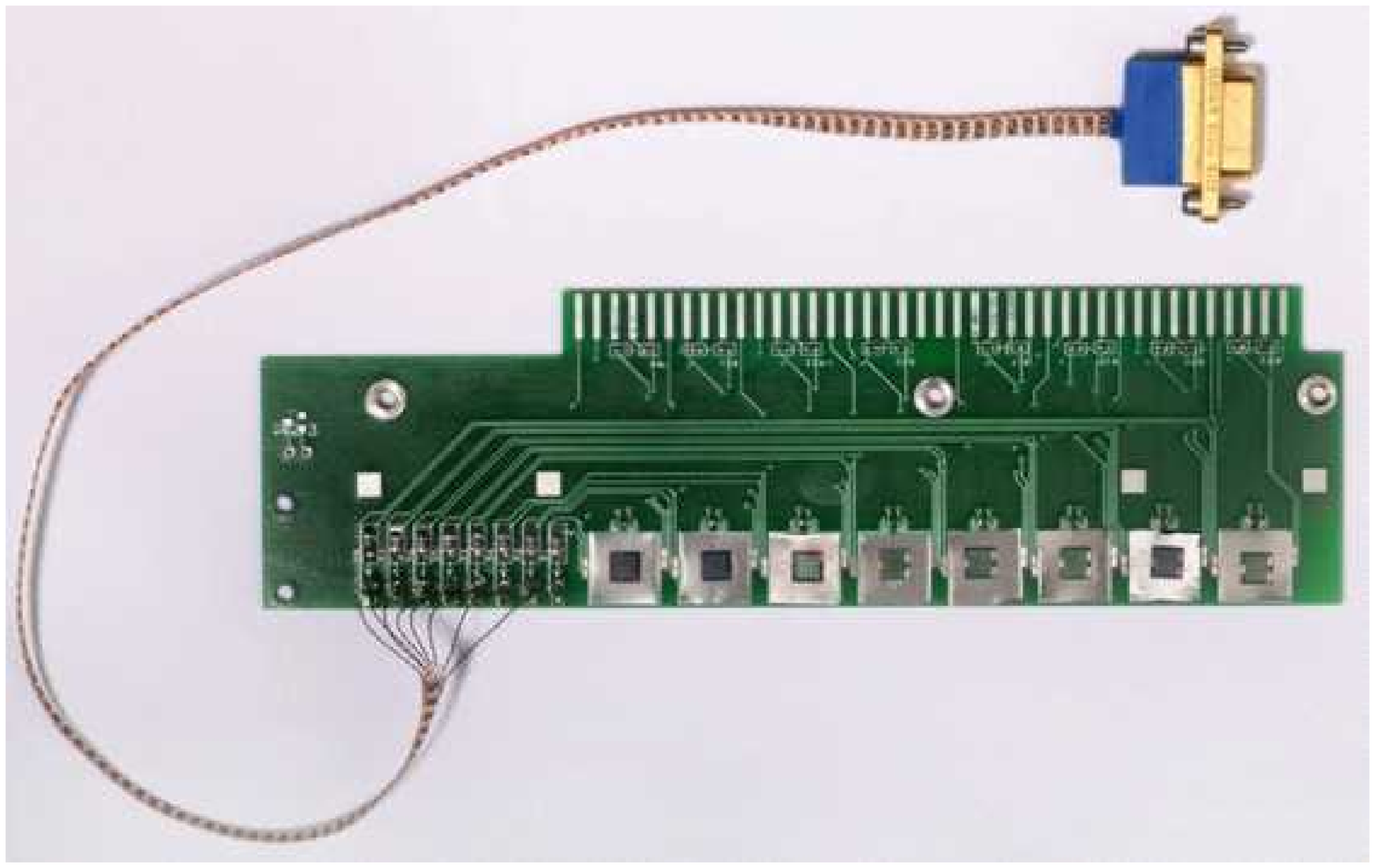}}
\caption[]{ A printed circuit board with eight 100-element squid arrays 
mounted individually on top of Niobium pads. 
The board is mounted inside the Cryoperm shield
and the Niobium films pin residual magnetic fields.
}
\label{fig:squids}
\end{minipage}
\hfill
\begin{minipage}{.475\textwidth}
\centerline{\includegraphics[height=2in]{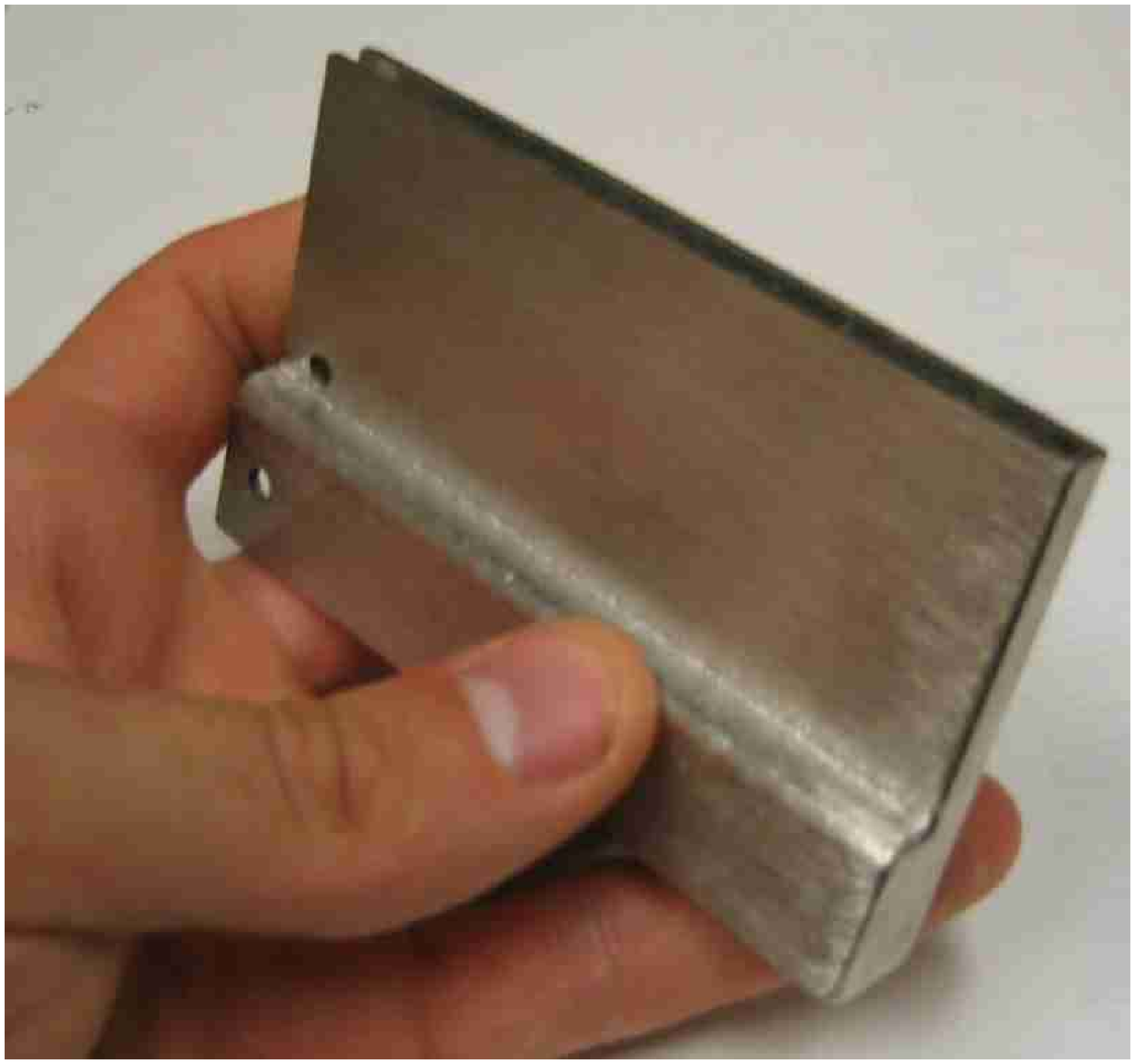}}
\caption[]{A prototype cryoperm shield is shown. The shield attenuates
  magnetic fields and provides mechanical support for the pc-board
  mounted SQUIDs.
}
\label{fig:cryoperm}
\end{minipage}
\end{figure}

The SQUID parameters must be chosen carefully to achieve the required noise performance
and dynamic range. 
It is necessary to use SQUID devices with a small input coil to
accommodate large amplitude carriers and reduce susceptibility to 
spurious pickup that degrades SQUID performance. This necessitates a large 
transimpedance of the SQUID amplifier to override the noise of the warm electronics. 
To meet these requirements, we utilize 100-element series array SQUID amplifiers 
supplied by NIST \cite{NIST-arrays}. 
The SQUIDs are packaged on pc-boards in groups of 8 as shown in Figure~\ref{fig:squids}.  
SQUIDs are extremely sensitive to magnetic fields and must be shielded very carefully;
each board is enclosed in a cryoperm shield as shown in Figure~\ref{fig:cryoperm}.
In addition, each SQUID array is individually mounted on a thin niobium film to pin 
the residual magnetic field. The attenuation achieved with this cryoperm/Nb shielding 
is better than a part-per-million at the frequencies of interest. 

Maintaining constant voltage bias for the bolometers necessitates that all impedances 
between the bolometer and the
bias resistor are much smaller than the bolometer resistance. 
Thus, the SQUID amplifier is operated with shunt feedback to obtain a low input 
impedance, while linearizing the SQUID response and extending the signal range to 
accommodate the high level bias carrier signals. 
The required gain-bandwidth product of the feedback loop requires short connections, 
so the SQUID controller is mounted directly on the side of the receiver cryostat. 
The noise of the readout system is less than 10~pA$/\sqrt{\rm{Hz}}$ and lies well below 
the noise floor of the detectors planned for SPT. 
The system has the large dynamic range and bandwidth (1 MHz) required for 
frequency domain multiplexing. 
The SQUID controller and demodulator 
are computer controlled with extensive monitoring and diagnostic capabilities.

The amplified comb of carriers are passed from the SQUID controller to the 
oscillator/demodulator boards shown in Figure~\ref{fig:oscboards}. 
Each board combines 16 demodulator channels. 
Since one demodulator channel is used for each bolometer, about 60 boards are 
required for SPT. 
Three 9U VME crates accommodate the readout, which is mounted in a separate rack 
near the cryostat. 
The high frequency bias carriers are generated by Direct Digital Synthesizers (DDS), 
which provide precise frequency and amplitude control with very low sideband noise. 
The same DDS that generates the bolometer bias also provides the local oscillator 
signal for the corresponding demodulator. The demodulator circuit utilizes a sampling 
demodulator with very high dynamic range ~\cite{tayloePatent} followed by an 8 pole 
low-pass anti-aliasing filter, and a 14 bit analog-to-digital converter.
A field programmable gate array assembles the data from all channels on a given board 
and streams it to the data acquisition computer. 
Final testing of the oscillator/demodulator production prototypes is underway.

\section{Science Goals and Implications for Observing Strategy} 
\label{sec:science}

\label{sec:clusters}
\begin{figure}[bt]
\begin{minipage}{.45\textwidth}
\includegraphics[width=2.8in]{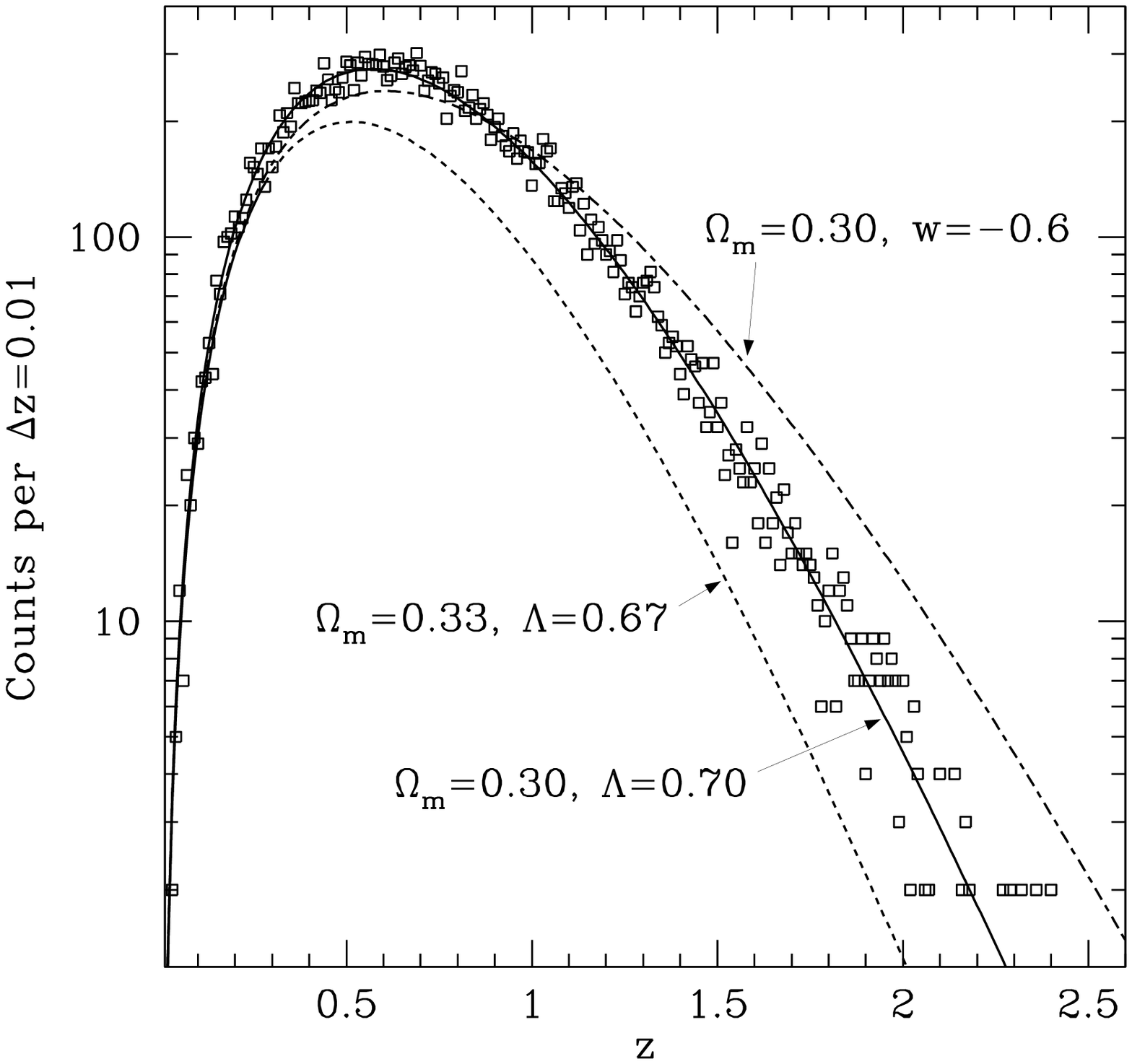}
\caption[]{
An illustration of the effect of cosmology on the expected
number of SZE detected galaxy clusters as a function of redshift.  The data 
points are appropriate for a 4000 square degree SPT survey with idealized
sensitivity.
The data points and the line passing through them were generated assuming
a canonical $\Omega_M = 0.3$, $\Omega_\Lambda = 0.7, \sigma_8 =1$ cosmology.
The other two lines show the large effect in the expected cluster counts
due to slight changes in the cosmology. The value of $\sigma_8$
was adjusted to give the same normalization for
the local cluster abundance in each model.
The bottom curve 
is for a model with more matter and correspondingly less dark energy.
The top curve at 
shows the effect of only a change in the equation of state
of the dark energy in the canonical model.
(Figure courtesy of G.\ Holder)
}
\label{fig:dndz}
\end{minipage}
\hfill
\begin{minipage}{.45\textwidth}
\includegraphics[width=3.0in,bb=50 373 312 686]{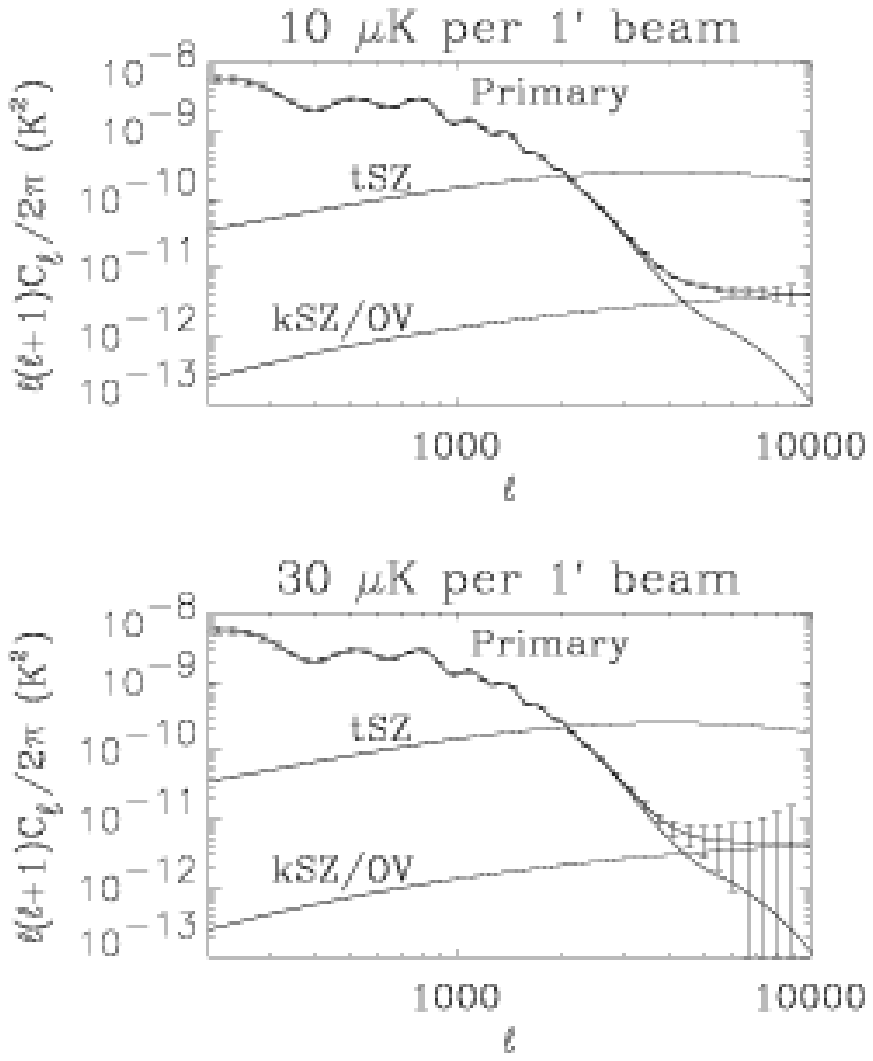}
\caption[]{An illustration of the potential of the SPT to measure 
fine-scale CMB anisotropy.  The two panels show statistical errors 
on the high-$\ell$ CMB power spectrum from 500 \sqdeg~of sky 
measured at two different levels of noise per $1'$ beam.  Both panels 
assume perfect subtraction of the thermal SZE signal and other 
astrophysical contaminants;  achieving the required accuracy
in this subtraction will be a significant challenge.  
(Spectra courtesy of W.\ Hu.)}
\label{fig:clcmb}
\end{minipage}
\vspace{0.1in}
\end{figure}

The high sensitivity and high angular resolution of the SPT will 
enable several ambitious scientific programs.  
The initial observational program 
will be a large survey for galaxy clusters detected by
the Sunyaev-Zel'dovich Effect (SZE).  The SZE is produced when CMB
photons scatter off the hot electron gas in galaxy 
clusters~\cite{sunyaev72}; the brightness of the SZE is 
nearly independent
of distance to the cluster, making it an ideal tool for conducting a
mass-limited cluster survey~\cite{carlstrom02}.  
The 
abundance of massive clusters as a function of redshift is highly
sensitive to the efficiency with which structure can grow, which is in
turn sensitive to the expansion history of the universe.  Therefore 
as shown in Figure~\ref{fig:dndz}, the redshift evolution of the
abundance of massive clusters
is critically sensitive to cosmological parameters 
such as the amount of dark energy
and its equation of state.
The SPT SZE survey will enable strong constraints on the amount
of and nature of dark energy in the universe.

\label{sec:cmb}
As reviewed in Section~\ref{sec:intro}, recent results from measurements of degree-scale anisotropies in the CMB
have spectacularly
confirmed predictions of the Hot Big Bang cosmological model, and
made precise measurements of many cosmological parameters.
New experiments are now focusing on characterizing the
temperature fluctuations on finer angular scales, where
secondary anisotropies are expected to dominate over fluctuations
imprinted on the last scattering surface~\cite{hu97b}.  

As shown in Figure~\ref{fig:clcmb}, the largest source of anisotropy
at multipole values $\ell > 2000$ is expected to be the thermal SZE.
Measurements of the angular power spectrum of this signal --
including or removing the massive clusters detected in the SZE survey
-- will allow tight determinations of the parameters $\sigma_8$ and $\Omega_M$
that are complementary to those obtained with analysis of the cluster
survey~\cite{komatsu02}.

The thermal SZE signal has a unique spectral signature.  There is a
null in the spectrum near 220~GHz, and the signal appears as a flux
decrement (relative to the 2.7K background) at frequencies below this
null and an increment at higher frequencies.  This opens the potential
for separating the SZE component from other contributions to the
CMB power spectrum, such as the kinetic SZE (KSZ, due to the net motion of
a cluster along the line of sight) and the Ostriker-Vishniac (OV)
effect, which is a similar effect produced by structures that are still in 
the linear regime.
Figure~\ref{fig:clcmb} shows how well an ideal 500 \sqdeg~SPT survey
could measure the fine-scale KSZ/OV anisotropy signal, assuming
perfect compensation for the thermal SZE signal and other astrophysical
contaminants.

\subsection{Atmospheric Noise}
\label{sec:atmo}

As discussed in Section~\ref{sec:site}, spatial fluctuations in atmospheric 
emission that are driven through the telescope beam by wind or 
scanning cause 
variations in the detector timestreams loosely referred to as 
``atmospheric noise".  The extent to which atmospheric noise dominates
instrument noise depends on the details of the site, telescope and
detector array properties, frequency bands
and observing strategy.

We have used a model of the fluctuations in atmospheric emission at the South Pole 
based on on the ACBAR observations~\cite{bussman04} to simulate two simple methods
of removing atmospheric noise for a variety of telescope scan 
speeds.  The first method is to apply a simple high-pass filter to the 
detector time-stream data, removing low-frequency atmospheric (and cosmic) 
signals.  The second method exploits the extensive overlap of beams 
from various 
detectors in the array at the height of the 
turbulent layer in the atmosphere~\cite{lay00,bussman04}.  
In practice,
we fit several spatial Fourier components across the array at
each time step to remove the common-mode signal from each detector's 
time-stream data.

\begin{figure}
\begin{center}
\includegraphics[width=3.0in,bb=55 360 415 720]{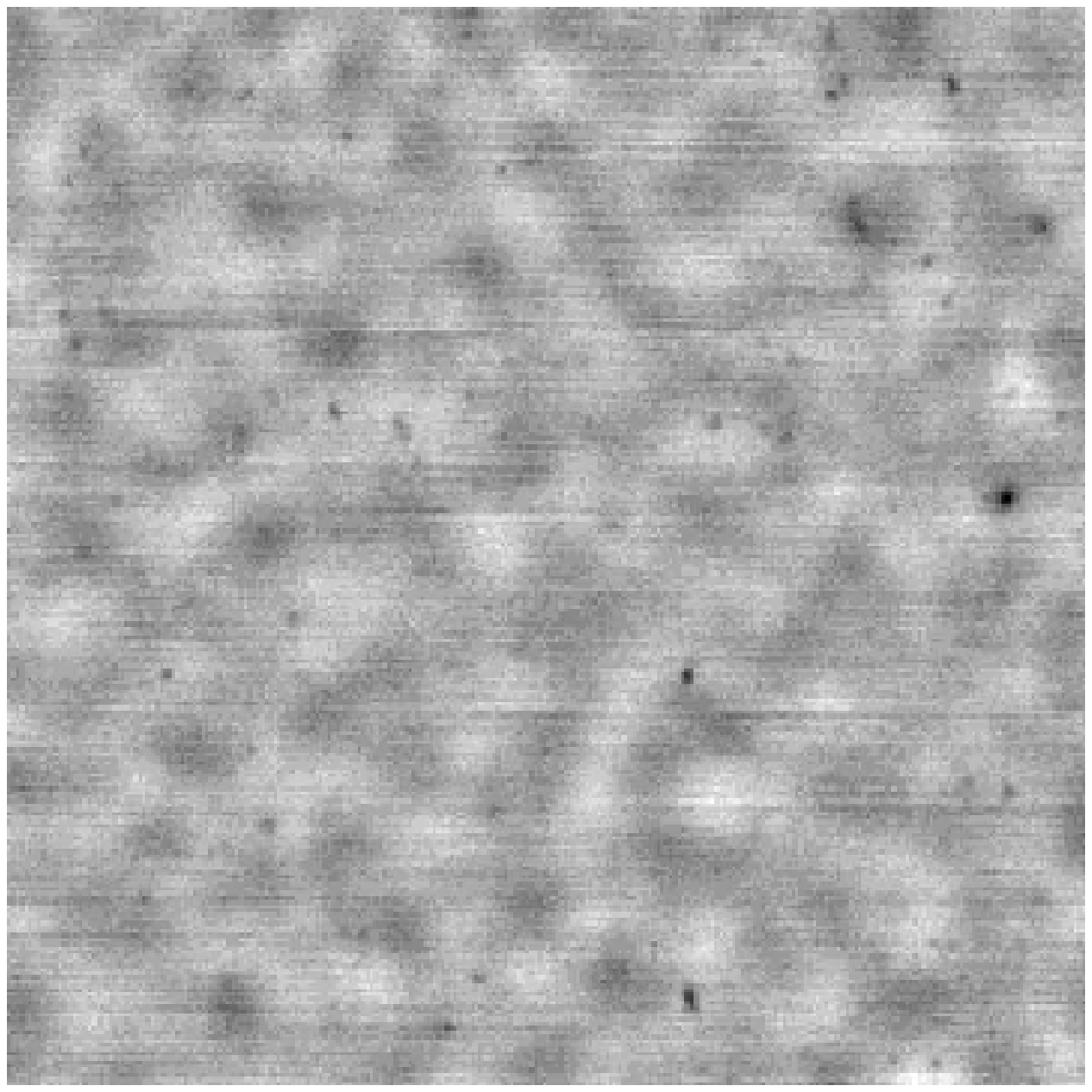}
\includegraphics[width=3.0in,bb=55 360 415 720]{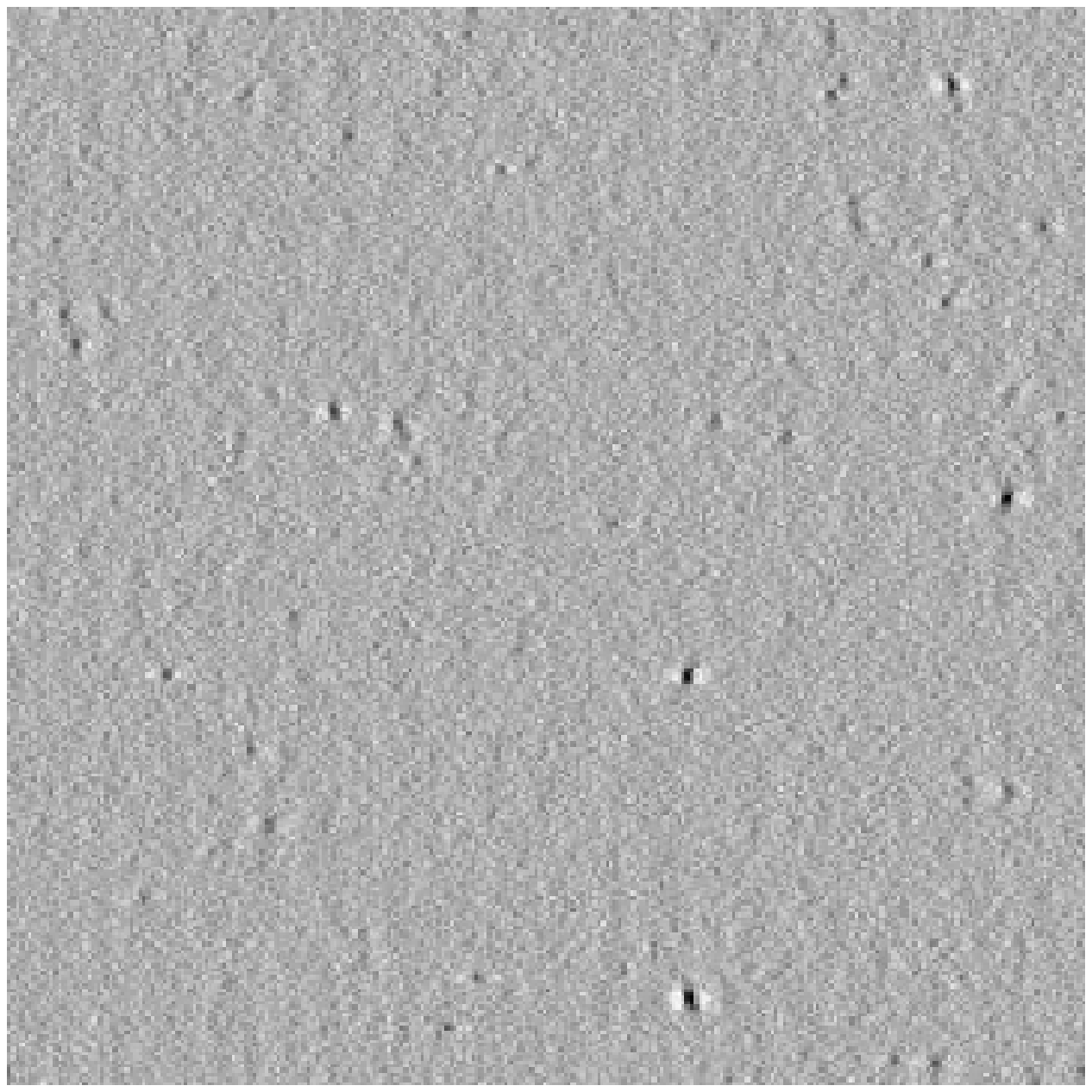}
\end{center}
  \caption{Two simple binned maps made from simulated 
array observations at 150 GHz
and a scan speed of 2$'s^{-1}$ of a CMB-plus-SZE sky with atmospheric
contamination and instrument noise with a $1/f$ knee at 0.1 Hz
(SZE maps courtesy of M. White~\cite{schulz03}).  The
images are 3 degrees across.  The two simulated observations were
filtered using different methods to achieve the same level of
atmospheric rejection.  
{\bf Left Panel:} results from a simulated observations in which the projection onto
16 low-frequency spatial modes across the array were subtracted at
each time sample.  The CMB structure is evident, as well as the 
cluster signals.  The horizontal striping due to $1/f$ noise could 
be removed by a very weak high-pass filter with negligible effect on the
CMB.
{\bf Right Panel:} results from a simulated observation in which
each individual bolometer time-stream data was high-passed at $\sim 0.3 \;
\mathrm{Hz}$.  This removes nearly all of the CMB structure, but leaves
the clusters quite visible.}
  \label{fig:maps}
\end{figure}

Figure~\ref{fig:maps}
shows results from simple binning of a simulated SPT observation of primary CMB
and SZE signals 
using a relatively slow scan rate of $2' s^{-1}$ with the 
data processed according to the two methods described above to remove atmospheric
fluctuations.
We have tuned both methods to remove the same amount of
atmospheric contamination.  It is clear in the figure that the
common-mode analysis retains cluster information on scales
beyond typical cluster radii and also retains much of the CMB information.
The high-pass filter, however, removed much of the larger-scale
information.  Galaxy clusters are easy to detect in both cases.  With these
algorithms and anticipated levels of detector and atmospheric noise, 
the cluster mass detection limit is only $10-20 \%$ higher at a scan 
rate of $2' s^{-1}$ compared with an ``infinite" scan speed.  Thus we 
find that even relatively slow scanning of the entire telescope is a viable 
observing strategy, enabled by the extremely stable atmospheric 
conditions at the South Pole.

\subsection{Frequency Band Considerations}
\label{sec:bandcalc}

Our choice of bands, integration time, and survey area
will be driven by particular science goals and the 
expected contaminants.  We have two main goals in this regard: 
\begin{enumerate}
\item Minimize the effect of primary CMB and other 
astrophysical signals on SZE cluster
detection and characterization, 
\item Construct a CMB map free of thermal SZE emission 
(and other astrophysical contaminants),
enabling a measure of the CMB power spectrum (and higher order statistics) and KSZ/OV effects 
at high $\ell$.
\end{enumerate}

Each of these goals will require broad spectral coverage;
atmospheric opacity
limits our choices of bands to narrow windows centered near 95, 150,
220, 270, and 350 GHz.  The expected sensitivity in each band 
(for single-moded, unpolarized detectors) is given in 
Table~\ref{tab:sensitivity}.
We are currently working with simulations to determine the most favorable
combination of bands and integration times (per band);  the results of 
these simulations are in large part driven by the assumptions
about the foreground emission. 

At the angular scales of galaxy clusters and
secondary CMB anisotropies, the dominant contaminants are expected to 
be extragalactic point sources of two flavors:  flat spectrum
radio sources, and luminous dusty galaxies.  With the $1'$ angular  
resolution of the SPT, these are difficult to separate from 
most cluster SZE signals simply by spatial filtering, so we turn to 
estimating the spectral signatures of these sources. 

Synchrotron emission from galaxies and AGNs is thought to be the main
mechanism for radio point source emission at frequencies less than $\sim 30$ GHz.  The steeply
falling spectrum of synchrotron emission should make these sources a
negligible contaminant at 150~GHz; however, there is evidence for a
population of flat or inverted-spectrum
sources~\cite{andrew70,trushkin03}.  Furthermore, there is strong
evidence that radio sources are preferentially associated with
clusters, thus amplifying their
contamination of SZE cluster surveys~\cite{pilkington64,cooray98}.

The level of the radio source contamination 
depends strongly on the number of such flat spectrum 
sources (as a function of flux) and their correlation with clusters.
Neither of these is yet known with enough certainty to make accurate
predictions.  Favorable estimates for the source density~\cite{knox03} 
and cluster correlation~\cite{cooray98} lead to the
conclusion that both the SZE survey and the CMB measurement will 
be unaffected by these sources.
However, even the most unfavorable estimates of the source
density~\cite{white04} and cluster correlation~\cite{massardi04} indicate
that only a small fraction of clusters will house a radio point source
that could seriously impede SZE detection at 150 GHz.  We have simulated
observations assuming such
worst-case radio point source scenarios, and find that
the contamination affects the minimum mass of 
clusters detected by only $10-20 \%$. Our understanding of 
the contamination from radio point sources will be aided with new
interferometric data at 100 GHz from instruments such as
the SZA\footnote{see http://astro.uchicago.edu/sza}.

Infrared point sources are a more likely contaminant of our survey.
The SCUBA and MAMBO instruments~\cite{smail97,kreysa98} have detected a
family of sources at 240, 350, and 670 GHz, 
thought to be dust-shrouded starburst galaxies that 
emit longward of 100 \mum \ ($\nu < 3000$GHz) as modified blackbodies.
Large numbers of detections over a range of fluxes have resulted in
reasonable statistics on the sources counts versus flux at 350 GHz.  Recently,
rest-frame spectral energy distributions and redshifts have been
determined for 73 of these objects~\cite{blain04}, providing a means to
extrapolate the 350 GHz counts into our other candidate observing
bands.  The extrapolations indicate that SPT observations at 150~GHz
will be contaminated by these sources.  For this reason
the SPT will have a high frequency ``dust" channel that will
be used as a monitor for these sources.  The exact band or combination of
bands that the SPT will use for this purpose is currently under study
using a combination of Fisher-matrix-type sensitivity
calculations and simulated observations.

\vskip 12pt

While much work remains, the South Pole Telescope and bolometer
array are 
currently on track for deployment to the NSF Amunsen-Scott South
Pole station
in late 2006. The SZE and CMB temperature anisotropy 
observations are scheduled to start at the beginning of
the 2007 Austral winter. 

%%%%%%%%%%%%%%%%%%%%%%%%%%%%%%%%%%%%%%%%%%%%%%%%%%%%%%%%%%%%%
\acknowledgments     %>>>> equivalent to \section*{ACKNOWLEDGMENTS}       
 
The SPT is supported by the U.S.A.\ National
Science Foundation under Grant No.\ OPP-0130612.
Additional support is provided by NSF Grant No.\ NSF PHY-0114422.
Work at LBNL is supported by the Director, Office of Science, Office of High Energy and
Nuclear Physics, of the U.S. Department of Energy under Contract No. DE-AC03-76SF00098. We thank VertexRSI and Raytheon Polar Support Corporation for
their support of the project and Eric Chauvin for valuable contributions to the 
telescope design.

%%%%%%%%%%%%%%%%%%%%%%%%%%%%%%%%%%%%%%%%%%%%%%%%%%%%%%%%%%%%%
%%%%% References %%%%%

%\bibliographystyle{spiebib}   %>>>> makes bibtex use spiebib.bst


\begin{thebibliography}{10}

\bibitem{penzias65}
A.~A. {Penzias} and R.~W. {Wilson}, ``{A Measurement of Excess Antenna
  Temperature at 4080 Mc/s.},'' {\em \apj} {\bf 142}, pp.~419--421, July 1965.

\bibitem{smoot92}
G.~F. Smoot {\em et~al.}, ``{Structure in the COBE Differential Microwave
  Radiometer First-Year Maps},'' {\em \apj} {\bf 396}, pp.~L1--L5, 1992.

\bibitem{bennett96}
C.~L. {Bennett}, A.~J. {Banday}, K.~M. {Gorski}, G.~{Hinshaw}, P.~{Jackson},
  P.~{Keegstra}, A.~{Kogut}, G.~F. {Smoot}, D.~T. {Wilkinson}, and E.~L.
  {Wright}, ``Four-year cobe dmr cosmic microwave background observations: Maps
  and basic results,'' {\em \apjl} {\bf 464}, p.~L1, June 1996.

\bibitem{mather94}
J.~C. {Mather}, E.~S. {Cheng}, D.~A. {Cottingham}, R.~E. {Eplee}, D.~J.
  {Fixsen}, T.~{Hewagama}, R.~B. {Isaacman}, K.~A. {Jensen}, S.~S. {Meyer},
  P.~D. {Noerdlinger}, S.~M. {Read}, L.~P. {Rosen}, R.~A. {Shafer}, E.~L.
  {Wright}, C.~L. {Bennett}, N.~W. {Boggess}, M.~G. {Hauser}, T.~{Kelsall},
  S.~H. {Moseley}, R.~F. {Silverberg}, G.~F. {Smoot}, R.~{Weiss}, and D.~T.
  {Wilkinson}, ``{Measurement of the cosmic microwave background spectrum by
  the COBE FIRAS instrument},'' {\em \apj} {\bf 420}, pp.~439--444, Jan. 1994.

\bibitem{fixsen96}
D.~J. {Fixsen}, E.~S. {Cheng}, J.~M. {Gales}, J.~C. {Mather}, R.~A. {Shafer},
  and E.~L. {Wright}, ``{The Cosmic Microwave Background Spectrum from the Full
  COBE FIRAS Data Set},'' {\em \apj} {\bf 473}, pp.~576--587, Dec. 1996.

\bibitem{miller99}
A.~D. {Miller}, R.~{Caldwell}, M.~J. {Devlin}, W.~B. {Dorwart}, T.~{Herbig},
  M.~R. {Nolta}, L.~A. {Page}, J.~{Puchalla}, E.~{Torbet}, and H.~T. {Tran},
  ``{A Measurement of the Angular Power Spectrum of the Cosmic Microwave
  Background from l = 100 to 400},'' {\em \apjl} {\bf 524}, pp.~L1--L4, Oct.
  1999.
\newblock astro-ph/9906421.

\bibitem{mauskopf00a}
P.~D. {Mauskopf}, P.~A.~R. {Ade}, P.~{de Bernardis}, J.~J. {Bock},
  J.~{Borrill}, A.~{Boscaleri}, B.~P. {Crill}, G.~{DeGasperis}, G.~{De Troia},
  P.~{Farese}, P.~G. {Ferreira}, K.~{Ganga}, M.~{Giacometti}, S.~{Hanany},
  V.~V. {Hristov}, A.~{Iacoangeli}, A.~H. {Jaffe}, A.~E. {Lange}, A.~T. {Lee},
  S.~{Masi}, A.~{Melchiorri}, F.~{Melchiorri}, L.~{Miglio}, T.~{Montroy}, C.~B.
  {Netterfield}, E.~{Pascale}, F.~{Piacentini}, P.~L. {Richards}, G.~{Romeo},
  J.~E. {Ruhl}, E.~{Scannapieco}, F.~{Scaramuzzi}, R.~{Stompor}, and
  N.~{Vittorio}, ``{Measurement of a Peak in the Cosmic Microwave Background
  Power Spectrum from the North American Test Flight of Boomerang},'' {\em
  \apjl} {\bf 536}, pp.~L59--L62, June 2000.

\bibitem{debernardis00}
P.~{de Bernardis}, P.~A.~R. {Ade}, J.~J. {Bock}, J.~R. {Bond}, J.~{Borrill},
  A.~{Boscaleri}, K.~{Coble}, B.~P. {Crill}, G.~{De Gasperis}, P.~C. {Farese},
  P.~G. {Ferreira}, K.~{Ganga}, M.~{Giacometti}, E.~{Hivon}, V.~V. {Hristov},
  A.~{Iacoangeli}, A.~H. {Jaffe}, A.~E. {Lange}, L.~{Martinis}, S.~{Masi},
  P.~V. {Mason}, P.~D. {Mauskopf}, A.~{Melchiorri}, L.~{Miglio}, T.~{Montroy},
  C.~B. {Netterfield}, E.~{Pascale}, F.~{Piacentini}, D.~{Pogosyan},
  S.~{Prunet}, S.~{Rao}, G.~{Romeo}, J.~E. {Ruhl}, F.~{Scaramuzzi},
  D.~{Sforna}, and N.~{Vittorio}, ``{A Flat Universe from High-Resolution Maps
  of the Cosmic Microwave Background Radiation},'' {\em \nat} {\bf 404},
  pp.~955--959, Apr. 2000.
\newblock astro-ph/0004404.

\bibitem{hanany00}
S.~{Hanany}, P.~{Ade}, A.~{Balbi}, J.~{Bock}, J.~{Borrill}, A.~{Boscaleri},
  P.~{de Bernardis}, P.~G. {Ferreira}, V.~V. {Hristov}, A.~H. {Jaffe}, A.~E.
  {Lange}, A.~T. {Lee}, P.~D. {Mauskopf}, C.~B. {Netterfield}, S.~{Oh},
  E.~{Pascale}, B.~{Rabii}, P.~L. {Richards}, G.~F. {Smoot}, R.~{Stompor},
  C.~D. {Winant}, and J.~H.~P. {Wu}, ``Maxima-1: A measurement of the cosmic
  microwave background anisotropy on angular scales of 10'-5$^\circ$,'' {\em
  \apjl} {\bf 545}, pp.~L5--L9, Dec. 2000.
\newblock astro-ph/0005123.

\bibitem{halverson02}
N.~W. {Halverson}, E.~M. {Leitch}, C.~{Pryke}, J.~{Kovac}, J.~E. {Carlstrom},
  W.~L. {Holzapfel}, M.~{Dragovan}, J.~K. {Cartwright}, B.~S. {Mason},
  S.~{Padin}, T.~J. {Pearson}, A.~C.~S. {Readhead}, and M.~C. {Shepherd},
  ``{Degree Angular Scale Interferometer First Results: A Measurement of the
  Cosmic Microwave Background Angular Power Spectrum},'' {\em \apj} {\bf 568},
  pp.~38--45, Mar. 2002.
\newblock astro-ph/0104489.

\bibitem{netterfield02}
C.~B. {Netterfield}, P.~A.~R. {Ade}, J.~J. {Bock}, J.~R. {Bond}, J.~{Borrill},
  A.~{Boscaleri}, K.~{Coble}, C.~R. {Contaldi}, B.~P. {Crill}, P.~{de
  Bernardis}, P.~{Farese}, K.~{Ganga}, M.~{Giacometti}, E.~{Hivon}, V.~V.
  {Hristov}, A.~{Iacoangeli}, A.~H. {Jaffe}, W.~C. {Jones}, A.~E. {Lange},
  L.~{Martinis}, S.~{Masi}, P.~{Mason}, P.~D. {Mauskopf}, A.~{Melchiorri},
  T.~{Montroy}, E.~{Pascale}, F.~{Piacentini}, D.~{Pogosyan}, F.~{Pongetti},
  S.~{Prunet}, G.~{Romeo}, J.~E. {Ruhl}, and F.~{Scaramuzzi}, ``{A Measurement
  by BOOMERANG of Multiple Peaks in the Angular Power Spectrum of the Cosmic
  Microwave Background},'' {\em \apj} {\bf 571}, pp.~604--614, June 2002.
\newblock astro-ph/0104460.

\bibitem{perlmutter99}
S.~Perlmutter, G.~Aldering, G.~Goldhaber, R.~Knop, P.~Nugent, P.~Castro,
  S.~Deustua, S.~Fabbro, A.~Goobar, D.~E. Groom, I.~M. Hook, A.~G. Kim, M.~Kim,
  J.~Lee, N.~Nunes, R.~Pain, C.~Pennypacker, R.~Quimby, C.~Lidman, R.~Ellis,
  M.~Irwin, R.~McMahon, P.~Ruiz-Lapuente, N.~Walton, B.~Schaefer, B.~Boyle,
  A.~Filippenko, T.~Matheson, A.~Fruchter, N.~Panagia, H.~J.~M. Newberg, and
  W.~Couch, ``Measurements of omega and lambda from 42 high-redshift
  supernovae,'' {\em \apj} {\bf 517}, p.~565, 1999.

\bibitem{reiss98}
A.~G. {Riess}, A.~V. {Filippenko}, P.~{Challis}, A.~{Clocchiatti},
  A.~{Diercks}, P.~M. {Garnavich}, R.~L. {Gilliland}, C.~J. {Hogan}, S.~{Jha},
  R.~P. {Kirshner}, B.~{Leibundgut}, M.~M. {Phillips}, D.~{Reiss}, B.~P.
  {Schmidt}, R.~A. {Schommer}, R.~C. {Smith}, J.~{Spyromilio}, C.~{Stubbs},
  N.~B. {Suntzeff}, and J.~{Tonry}, ``Observational evidence from supernovae
  for an accelerating universe and a cosmological constant,'' {\em \aj} {\bf
  116}, pp.~1009--1038, Sept. 1998.

\bibitem{bennett03}
C.~L. {Bennett}, M.~{Halpern}, G.~{Hinshaw}, N.~{Jarosik}, A.~{Kogut},
  M.~{Limon}, S.~S. {Meyer}, L.~{Page}, D.~N. {Spergel}, G.~S. {Tucker},
  E.~{Wollack}, E.~L. {Wright}, C.~{Barnes}, M.~R. {Greason}, R.~S. {Hill},
  E.~{Komatsu}, M.~R. {Nolta}, N.~{Odegard}, H.~V. {Peiris}, L.~{Verde}, and
  J.~L. {Weiland}, ``{First-Year Wilkinson Microwave Anisotropy Probe (WMAP)
  Observations: Preliminary Maps and Basic Results},'' {\em \apjs} {\bf 148},
  pp.~1--27, Sept. 2003.
\newblock astro-ph/0302207.

\bibitem{kuo04}
C.~L. {Kuo}, P.~A.~R. {Ade}, J.~J. {Bock}, C.~{Cantalupo}, M.~D. {Daub},
  J.~{Goldstein}, W.~L. {Holzapfel}, A.~E. {Lange}, M.~{Lueker}, M.~{Newcomb},
  J.~B. {Peterson}, J.~{Ruhl}, M.~C. {Runyan}, and E.~{Torbet},
  ``{High-Resolution Observations of the Cosmic Microwave Background Power
  Spectrum with ACBAR},'' {\em \apj} {\bf 600}, pp.~32--51, Jan. 2004.
\newblock astro-ph/0202289.

\bibitem{mason02}
B.~S. {Mason}, T.~J. {Pearson}, A.~C.~S. {Readhead}, M.~C. {Shepherd},
  J.~{Sievers}, P.~S. {Udomprasert}, J.~K. {Cartwright}, A.~J. {Farmer},
  S.~{Padin}, S.~T. {Myers}, J.~R. {Bond}, C.~R. {Contaldi}, U.~{Pen},
  S.~{Prunet}, D.~{Pogosyan}, J.~E. {Carlstrom}, J.~{Kovac}, E.~M. {Leitch},
  C.~{Pryke}, N.~W. {Halverson}, W.~L. {Holzapfel}, P.~{Altamirano},
  L.~{Bronfman}, S.~{Casassus}, J.~{May}, and M.~{Joy}, ``{The Anisotropy of
  the Microwave Background to l = 3500: Deep Field Observations with the Cosmic
  Background Imager},'' {\em \apj} {\bf 591}, pp.~540--555, July 2003.
\newblock astro-ph/0205384.

\bibitem{spergel03}
D.~N. {Spergel}, L.~{Verde}, H.~V. {Peiris}, E.~{Komatsu}, M.~R. {Nolta}, C.~L.
  {Bennett}, M.~{Halpern}, G.~{Hinshaw}, N.~{Jarosik}, A.~{Kogut}, M.~{Limon},
  S.~S. {Meyer}, L.~{Page}, G.~S. {Tucker}, J.~L. {Weiland}, E.~{Wollack}, and
  E.~L. {Wright}, ``{First Year Wilkinson Microwave Anisotropy Probe (WMAP)
  Observations: Determination of Cosmological Parameters},'' {\em ApJ
  submitted} , Feb. 2003.
\newblock astro-ph/0302209.

\bibitem{hu97}
W.~Hu and M.~White, ``The dampling tail of cosmic microwave background
  anisotropies,'' {\em \apj} {\bf 479}, p.~568, 1997.

\bibitem{sunyaev70}
R.~Sunyaev and Y.~Zel'dovich, ``The spectrum of primordial radiation, its
  distortions and their significance,'' {\em Comments Astrophys. Space Phys.}
  {\bf 2}, p.~66, 1970.

\bibitem{sunyaev72}
R.~A. {Sunyaev} and Y.~B. {Zeldovich}, ``{The Observations of Relic Radiation
  as a Test of the Nature of X-Ray Radiation from the Clusters of Galaxies},''
  {\em Comments on Astrophysics and Space Physics} {\bf 4}, pp.~173--+, Nov.
  1972.

\bibitem{carlstrom02}
J.~E. {Carlstrom}, G.~P. {Holder}, and E.~D. {Reese}, ``{Cosmology with the
  Sunyaev-Zel'dovich Effect},'' {\em \araa} {\bf 40}, pp.~643--680, 2002.

\bibitem{holder01b}
G.~{Holder}, Z.~{Haiman}, and J.~J. {Mohr}, ``{Constraints on $\Omega_M$,
  $\Omega_{\Lambda}$, and $\sigma_8$ from Galaxy Cluster Redshift
  Distributions},'' {\em \apjl} {\bf 560}, pp.~L111--L114, Oct. 2001.

\bibitem{haiman01}
Z.~{Haiman}, J.~J. {Mohr}, and G.~P. {Holder}, ``{Constraints on Cosmological
  Parameters from Future Galaxy Cluster Surveys},'' {\em \apj} {\bf 553},
  pp.~545--561, June 2001.

\bibitem{hu_w97}
W.~Hu and M.~White, ``A {CMB} polarization primer,'' {\em {New Astronomy}} {\bf
  2}, pp.~323--344, 1997.
\newblock astro-ph/9706147.

\bibitem{kovac02}
J.~M. Kovac, E.~M. Leitch, C.~Pryke, J.~E. Carlstrom, N.~W. Halverson, and
  W.~L. Holzapfel, ``Detection of polarization in the cosmic microwave
  background using {DASI},'' {\em Nature} {\bf 420}, p.~772, December 2002.
\newblock astro-ph/0209478.

\bibitem{kogut03a}
A.~{Kogut}, D.~N. {Spergel}, C.~{Barnes}, C.~L. {Bennett}, M.~{Halpern},
  G.~{Hinshaw}, N.~{Jarosik}, M.~{Limon}, S.~S. {Meyer}, L.~{Page}, G.~S.
  {Tucker}, E.~{Wollack}, and E.~L. {Wright}, ``{First-Year Wilkinson Microwave
  Anisotropy Probe (WMAP) Observations: Temperature-Polarization
  Correlation},'' {\em \apjs} {\bf 148}, pp.~161--173, Sept. 2003.
\newblock astro-ph/0302213.

\bibitem{polnarev85}
A.~G. {Polnarev}, ``{Polarization and Anisotropy Induced in the Microwave
  Background by Cosmological Gravitational Waves},'' {\em \sovast} {\bf 29},
  pp.~607--613, Dec. 1985.

\bibitem{crittenden93}
R.~{Crittenden}, R.~L. {Davis}, and P.~J. {Steinhardt}, ``{Polarization of the
  Microwave Background Due to Primordial Gravitational Waves},'' {\em \apjl}
  {\bf 417}, pp.~L13--L16, Nov. 1993.
\newblock astro-ph/9306027.

\bibitem{seljak97a}
U.~{Seljak}, ``{Measuring Polarization in the Cosmic Microwave Background},''
  {\em \apj} {\bf 482}, pp.~6--16, June 1997.
\newblock astro-ph/9608131.

\bibitem{kamionkowski97b}
M.~{Kamionkowski}, A.~{Kosowsky}, and A.~{Stebbins}, ``{A Probe of Primordial
  Gravity Waves and Vorticity},'' {\em Physical Review Letters} {\bf 78},
  pp.~2058--2061, Mar. 1997.
\newblock astro-ph/9609132.

\bibitem{seljak97}
U.~{Seljak} and M.~{Zaldarriaga}, ``{Signature of Gravity Waves in the
  Polarization of the Microwave Background},'' {\em Physical Review Letters}
  {\bf 78}, pp.~2054--2057, Mar. 1997.
\newblock astro-ph/9609169.

\bibitem{okamoto02}
T.~{Okamoto} and W.~{Hu}, ``{Cosmic microwave background lensing reconstruction
  on the full sky},'' {\em \prd} {\bf 67}, pp.~083002--+, Apr. 2003.
\newblock astro-ph/0301031.

\bibitem{knox02}
L.~{Knox} and Y.~{Song}, ``{A limit on the detectability of the energy scale of
  inflation},'' {\em Physical Review Letters} {\bf 89}, p.~011303 (4 pages),
  2002.
\newblock astro-ph/0202286.

\bibitem{schwerdtfeger}
W.~Schwerdtfeger, {\em Weather and Climate of the Antarctic}, Elsevier,
  Amsterdam, 1984.

\bibitem{chamberlin94}
R.~A. Chamberlin and J.~Bally, ``The 225 {GH}z opacity of the {S}outh {P}ole
  sky derived from continual radiometric measurements of the sky brightness
  temperature,'' {\em \ao} {\bf 33}, p.~1095, 1994.

\bibitem{chamberlin95}
R.~A. Chamberlin and J.~Bally, ``The observed relationship between the {S}outh
  {P}ole 225 {GH}z atmospheric opacity and the water vapor column density,''
  {\em Int. J. Infrared and Millimeter Waves} {\bf 16}, p.~907, 1995.

\bibitem{chamberlin97}
R.~A. Chamberlin, A.~P. Lane, and A.~A. Stark, ``The 492 {GH}z atmospheric
  opacity at the {G}eographic {S}outh {P}ole,'' {\em \apj} {\bf 476}, p.~428,
  1997.

\bibitem{lane98}
A.~P. Lane, ``Submillimeter transmission at {S}outh {P}ole,'' in {\em
  Astrophysics from {A}ntarctica},  G.~Novak and R.~H. Landsberg, eds., {\em
  ASP Conf. Ser. 141} {\bf 141}, p.~289, ASP, (San Francisco), 1998.

\bibitem{stark01}
A.~A. Stark, J.~Bally, S.~P. Balm, T.~M. Bania, A.~D. Bolatto, R.~A.
  Chamberlin, G.~Engargiola, M.~Huang, J.~G. Ingalls, K.~Jacobs, J.~M. Jackson,
  J.~W. Kooi, A.~P. Lane, K.-Y. Lo, R.~D. Marks, C.~L. Martin, D.~Mumma,
  R.~Ojha, R.~Schieder, J.~Staguhn, J.~Stutzki, C.~K. Walker, R.~W. Wilson,
  G.~A. Wright, X.~Zhang, P.~Zimmermann, and R.~Zimmermann, ``The {A}ntarctic
  {S}ubmillimeter {T}elescope and {R}emote {O}bservatory ({AST/RO}),'' {\em
  \pasp} {\bf 113}, p.~567, 2001.

\bibitem{peterson03}
J.~B. Peterson, S.~J.~E. Radford, P.~A.~R. Ade, R.~A. Chamberlin, M.~J.
  O'Kelly, K.~M. Peterson, and E.~Schartman, ``Stability of the submillimeter
  brightness of the atmosphere above {M}auna {K}ea, {C}hajnantor, and the
  {S}outh {P}ole,'' {\em \pasp} {\bf 115}, pp.~383--388, 2003.

\bibitem{chamberlin01}
R.~A. Chamberlin, ``{S}outh {P}ole submillimeter sky opacity and correlations
  with radiosonde observations,'' {\em J. Geophys. Res. Atmospheres} {\bf 106
  (D17)}, pp.~20101--20113, 2001.

\bibitem{lay98}
O.~P. {Lay} and N.~W. {Halverson}, ``The impact of atmospheric fluctuations on
  degree-scale imaging of the cosmic microwave background,'' {\em \apj} {\bf
  543}, pp.~787--798, Nov. 2000.

\bibitem{radford96}
S.~J.~E. Radford, G.~Reiland, and B.~Shillue, ``Site test interferometer,''
  {\em PASP} {\bf 108}, p.~441, 1996.

\bibitem{holdaway95}
M.~A. Holdaway, S.~J.~E. Radford, F.~N. Owen, and S.~M. Foster, ``Fast
  switching phase calibration: Effectiveness at {M}auna {K}ea and
  {C}hajnantor,'' Millimeter Array Technical Memo 139, NRAO, 1995.

\bibitem{runyan03a}
M.~C. {Runyan}, P.~A.~R. {Ade}, R.~S. {Bhatia}, J.~J. {Bock}, M.~D. {Daub},
  J.~H. {Goldstein}, C.~V. {Haynes}, W.~L. {Holzapfel}, C.~L. {Kuo}, A.~E.
  {Lange}, J.~{Leong}, M.~{Lueker}, M.~{Newcomb}, J.~B. {Peterson},
  C.~{Reichardt}, J.~{Ruhl}, G.~{Sirbi}, E.~{Torbet}, C.~{Tucker}, A.~D.
  {Turner}, and D.~{Woolsey}, ``{ACBAR: The Arcminute Cosmology Bolometer Array
  Receiver},'' {\em \apjs} {\bf 149}, pp.~265--287, Dec. 2003.
\newblock astro-ph/0303515.

\bibitem{bussman04}
M.~Bussman, C.~L. Kuo, and W.~L. Holzapfel, ``{Atmsopheric noise in
  mm-wavelenghth bands at the South Pole},'' {\em In Preparation} , 2004.

\bibitem{serabyn02}
E.~{Serabyn} and J.~{Pardo}, ``{FTS measurements of submillimeter
  quasi-continuum atmospheric opacity terms},'' in {\em ASP Conf. Ser. 266:
  Astronomical Site Evaluation in the Visible and Radio Range},  pp.~206--+,
  2002.

\bibitem{stark00}
A.~A. Stark, ``Design considerations for large detector arrays on
  submillimeter-wave telescopes,'' in {\em Radio Telescopes},  H.~R. Butcher,
  ed.,  {\bf 4015}, p.~434, July 2000.

\bibitem{stark03b}
A.~A. Stark, ``Meeting the optical requirements of large focal-plane arrays,''
  in {\em Proceedings of the Fourteenth Intern. Symp. on Space THz Technology},
   C.~Groppi, ed., 2003.

\bibitem{griffin02}
M.~J. Griffin, J.~J. Bock, and W.~K. Gear, ``The relative performance of filled
  and feedhorn-coupled focal plane architectures,'' {\em Applied Optics} {\bf
  41}(31), pp.~6543--6554, 2002.
\newblock astro-ph/0205264.

\bibitem{dragone82}
C.~Dragone, ``A first-order treatment of aberrations in {C}assegrainian and
  {G}regorian antennas,'' {\em IEEE Transactions on Antennas and Propagation}
  {\bf 30}(3), pp.~331--339, 1982.

\bibitem{Spieler02}
H.~Spieler, ``Frequency domain multiplexing for large scale bolometer arrays,''
  in {\em Monterey Far-IR, Sub-mm and mm Detector Technology Workshop
  proceedings},  J.~Wolf, J.~Farhoomand, and C.~McCreight, eds., pp.~243--249,
  2002.
\newblock NASA/CP-2003-21140 and LBNL-49993,
  http://www-library.lbl.gov/docs/LBNL/499/93/PDF/LBNL-49993.pdf.

\bibitem{Lanting04}
T.~M. Lanting, H.-M. Cho, J.~Clarke, M.~A. Dobbs, A.~T. Lee, M.~Lueker, P.~L.
  Richards, A.~D. Smith, and H.~G. Spieler, ``Frequency domain multiplexing for
  bolometer arrays,'' {\em Nuclear Instruments and Methods in Physics Research
  A} {\bf 520}, pp.~548--550, 2004.

\bibitem{NIST-arrays}
M.~Huber, P.~Neil, R.~Benson, D.~Burns, A.~Corey, C.~Flynn, Y.~Kitaygorodskaya,
  O.~Massihzadeh, J.~Martinis, and G.~Hilton, ``{DC SQUID} series array
  amplifiers with 120 {MHz} bandwidth (corrected),'' {\em IEEE Transactions on
  Applied Superconductivity} {\bf 11}, pp.~4048--4053, 2001.

\bibitem{tayloePatent}
D.~Tayloe, ``{T}ayloe {M}ixer,'' {\em US Patent} {\bf 6,230,000 B1}, 2001.

\bibitem{hu97b}
W.~Hu, N.~Sugiyama, and J.~Silk, ``The physics of microwave background
  anisotropies,'' {\em Nature} {\bf 386}, pp.~37--43, 1997.
\newblock astro-ph/9604166.

\bibitem{komatsu02}
E.~{Komatsu} and U.~{Seljak}, ``{The Sunyaev-Zel'dovich angular power spectrum
  as a probe of cosmological parameters},'' {\em \mnras} {\bf 336},
  pp.~1256--1270, Nov. 2002.
\newblock astro-ph/0205468.

\bibitem{lay00}
O.~P. {Lay} and N.~W. {Halverson}, ``{The Impact of Atmospheric Fluctuations on
  Degree-Scale Imaging of the Cosmic Microwave Background},'' {\em \apj} {\bf
  543}, pp.~787--798, Nov. 2000.
\newblock astro-ph/9905369.

\bibitem{schulz03}
A.~E. {Schulz} and M.~{White}, ``{Surveys of Galaxy Clusters with the
  Sunyaev-Zel'dovich Effect},'' {\em \apj} {\bf 586}, pp.~723--730, Apr. 2003.
\newblock astro-ph/0210667.

\bibitem{andrew70}
B.~H. {Andrew} and J.~D. {Kraus}, ``{Radio Sources with Flat Spectra},'' {\em
  \apjl} {\bf 159}, pp.~L45--L50, Jan. 1970.

\bibitem{trushkin03}
S.~{Trushkin}, ``Radio spectra of the {WMAP} catalog sources,'' {\em
  Bull.Spec.Astrophys.Obs.N.Caucasus} {\bf 55}, pp.~90--132, 2003.
\newblock astro-ph/0307205.

\bibitem{pilkington64}
J.~D.~H. {Pilkington}, ``{Radio sources and rich clusters of galaxies},'' {\em
  \mnras} {\bf 128}, pp.~103--+, 1964.

\bibitem{cooray98}
A.~R. {Cooray}, L.~{Grego}, W.~L. {Holzapfel}, M.~{Joy}, and J.~E. {Carlstrom},
  ``{Radio Sources in Galaxy Clusters at 28.5 GHz},'' {\em \aj} {\bf 115},
  p.~1388, Apr. 1998.
\newblock astro-ph/9711218.

\bibitem{knox03}
L.~{Knox}, G.~P. {Holder}, and S.~E. {Church}, ``Effects of sub-mm and radio
  point sources on the recovery of {S}unyaev-{Z}eldovich galaxy cluster
  parameters,''
\newblock astro-ph/0309643.

\bibitem{white04}
M.~{White} and S.~{Majumdar}, ``{Point Sources in the Context of Future SZ
  Surveys},'' {\em \apj} {\bf 602}, pp.~565--570, Feb. 2004.
\newblock astro-ph/0105229.

\bibitem{massardi04}
M.~{Massardi} and G.~{De Zotti}, ``Radio source contamination of the
  {S}unyaev-{Z}eldovich effect in galaxy clusters,'' {\em \aap} .
\newblock submitted, astro-ph/0405323.

\bibitem{smail97}
I.~{Smail}, R.~J. {Ivison}, and A.~W. {Blain}, ``A deep sub-millimeter survey
  of lensing clusters: A new window on galaxy formation and evolution,'' {\em
  \apjl} {\bf 490}, p.~L5, Nov. 1997.

\bibitem{kreysa98}
E.~{Kreysa}, H.~{Gemuend}, J.~{Gromke}, C.~G. {Haslam}, L.~{Reichertz}, E.~E.
  {Haller}, J.~W. {Beeman}, V.~{Hansen}, A.~{Sievers}, and R.~{Zylka},
  ``{Bolometer array development at the Max-Planck-Institut fuer
  Radioastronomie},'' in {\em Proc. SPIE Vol. 3357, p. 319-325, Advanced
  Technology MMW, Radio, and Terahertz Telescopes, Thomas G. Phillips; Ed.},
  pp.~319--325, July 1998.

\bibitem{blain04}
A.~W. {Blain}, S.~C. {Chapman}, I.~{Smail}, and R.~{Ivison}, ``Accurate {SED}s
  and selection effects for high-redshift dusty galaxies: a new hot population
  to discover with {S}pitzer?,'' {\em \apj} .
\newblock in press, astro-ph/0404438.

\end{thebibliography}
\end{document}